\documentclass[letterpaper,twocolumn,aps,prx,floatfix,superscriptaddress,noeprint]{revtex4-2}

\usepackage{natbib}
\usepackage{float}
\usepackage[caption=false]{subfig}
\usepackage[section]{placeins}
\usepackage{url}
\usepackage{amsmath}
\usepackage{mathtools}
\usepackage{enumitem}
\usepackage{amssymb}
\usepackage{dsfont}
\usepackage{breqn}
\usepackage{hyperref}

\usepackage{graphicx}

\usepackage{xcolor}
\definecolor{green_cust}{RGB}{0,154,85}
\definecolor{red_cust}{RGB}{173,49,54}
\definecolor{blue_cust}{RGB}{0,103,148}
\hypersetup{colorlinks=true, linkcolor=red_cust, citecolor=green_cust, filecolor=blue_cust, urlcolor=blue_cust}

\usepackage[capitalise]{cleveref}
\usepackage{gensymb}
\usepackage[euler]{textgreek}
\usepackage{lipsum}

\makeatletter
\let\cat@comma@active\@empty
\makeatother

\graphicspath{ {./figs/}}
\newcommand{\op}[1]{\boldsymbol #1}
\newcommand{\ket}[1]{|#1\rangle}
\newcommand{\bra}[1]{\langle #1|}

\usepackage{amsmath}

\newcommand*{\Figref}[1]{Fig.~\hyperref[#1]{\ref*{#1}}}

\newcommand{\RNum}[1]{\uppercase\expandafter{\romannumeral #1\relax}}
\newcommand{\circlenum}[1]{\raisebox{.5pt}{\textcircled{\raisebox{-.9pt} {#1}}}}

\begin{document}	
\title{Ultrastrong light-matter interaction in a multimode photonic crystal}
\date{\today}

\author{Andrei Vrajitoarea}
\email{andrei.v@nyu.edu}
\affiliation{Department of Electrical Engineering, Princeton University, Princeton, New Jersey 08540, USA
}

\author{Ron Belyansky}
\author{Rex Lundgren}
\author{Seth Whitsitt}
\affiliation{Joint Center for Quantum Information and Computer Science, NIST/University of Maryland, College Park, Maryland 20742 USA}
\affiliation{Joint Quantum Institute, NIST/University of Maryland, College Park, Maryland 20742 USA}

\author{Alexey V. Gorshkov}
\affiliation{Joint Center for Quantum Information and Computer Science, NIST/University of Maryland, College Park, Maryland 20742 USA}
\affiliation{Joint Quantum Institute, NIST/University of Maryland, College Park, Maryland 20742 USA}

\author{Andrew A. Houck}
\email{aahouck@princeton.edu}
\affiliation{Department of Electrical Engineering, Princeton University, Princeton, New Jersey 08540, USA
}

\begin{abstract}
Harnessing the interaction between light and matter at the quantum level has been a central theme in atomic physics and quantum optics, with applications from quantum computation to quantum metrology. Combining complex interactions with photonic synthetic materials provides an opportunity to investigate novel quantum phases and phenomena, establishing interesting connections to condensed matter physics. Here we explore many-body phenomena with a single artificial atom coupled to the many discrete modes of a photonic crystal. This experiment reaches the ultrastrong light-matter coupling regime using the circuit quantum electrodynamics paradigm, by galvanically coupling a highly nonlinear fluxonium qubit to a tight-binding lattice of microwave resonators. In this regime, the transport of a single photon becomes a many-body problem, owing to the strong participation of multi-photon bound states arising from interactions that break particle number conservation. Exploiting the effective photon-photon interactions mediated by the qubit, the transport of multiple photons leads to complex multimode dynamics that can be employed for generating a continuous reservoir of strongly-correlated photons, an important resource for quantum networks. This work opens exciting prospects for exploring nonlinear quantum optics at the single-photon level and stabilizing entangled many-body phases of light.
\end{abstract}
\maketitle
\pagenumbering{arabic}

\section{Introduction}

\begin{figure*}[!t]
	\begin{center}
		\includegraphics[width=0.95\textwidth]{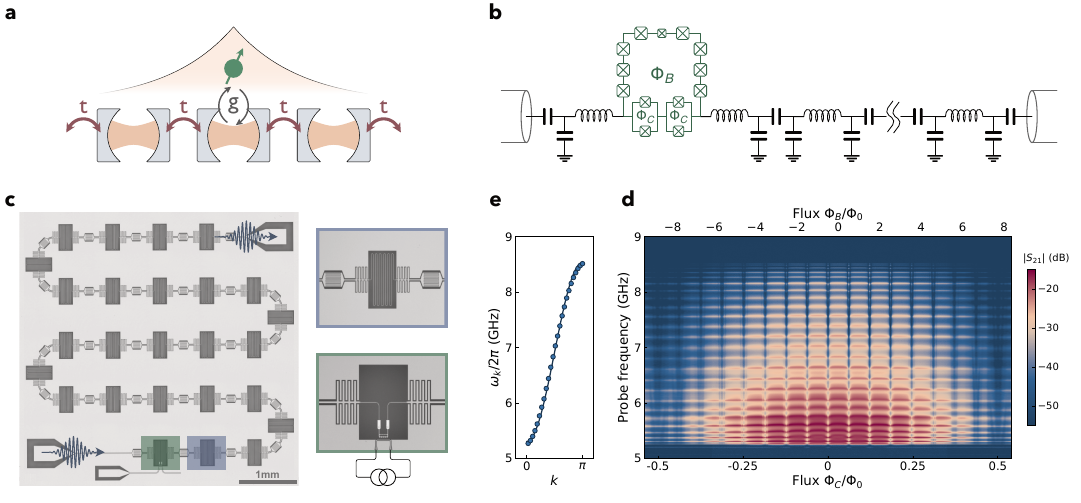}
	\end{center}
	\vspace{-0.5cm}
	\caption{
	\label{fig:1}\textbf{Photonic crystal platform.} 
	Atom-photon dressed bound states are explored in the waveguide quantum electrodynamics model illustrated in \textbf{a}. The physical system, shown in \textbf{b}, consists of a fluxonium artificial atom galvanically coupled to a tight-binding lattice of photons implemented as a lumped-element metamaterial. The fluxonium energy levels and its coupling inductance to the photonic crystal are tuned in situ with the externally applied magnetic flux $\Phi_{\text{B}}$ and $\Phi_{\text{C}}$, respectively. The photonic crystal device is patterned on a large-area superconducting circuit.
	As shown in the optical microscope image in \textbf{c}, each of $M = 26$ unit cells is defined by a microwave resonator (blue), and the fluxonium circuit is coupled to the edge resonator (red). \textbf{d.} The dressed eigenmodes of the metamaterial are spectroscopically probed in transmission as the qubit resonance and its coupling are tuned with the magnetic flux bias. \textbf{e.} The eigenmode frequencies, measured when the qubit is far detuned, match the theoretical energy dispersion $\omega_k = \omega_c - 2J\cos(k)$ (solid line) for a 1D tight-binding chain of cavities, with a resonance frequency $\omega_c$, coupled with a tunneling rate $J$. Here $k = n \pi/(M+1)$, with $n = 1, \dots, M$, labels the eigenmodes in the order of increasing energy. 
	} 
\end{figure*}

Manipulating and studying the behavior of lab-designed quantum systems composed of interacting particles, such as cold atoms~\cite{Bloch_NatPhys2012, Browaeys_NatPhys2020} or trapped ions~\cite{Blatt_NatPhys2012}, offers a unique opportunity for understanding fascinating phenomena in many-body physics. Employing quantum optical platforms for creating synthetic quantum materials has become increasingly popular,  expanding the experimental toolbox by leveraging the flexibility of the light-matter interface to extend interactions between atoms beyond the free-space limit using photonic structures with a tailored dispersion ~\cite{Chang_RMP2018}, or to mediate interactions between photons through a nonlinear atomic medium~\cite{Carusotto_NatPhys2020}.

One fascinating light-matter interface arises from the strong hybridization of an atom with the band structure of a photonic crystal. Such an interface creates unique atom-photon dressed states, where the photonic excitation is exponentially localized at the atom position~\cite{Bykov75, Sajeev_PRL1990}. The first detection of a single-photon bound state was demonstrated in the microwave domain, by coupling a superconducting qubit to a stepped impedance waveguide~\cite{Liu_NatPhys2017}.
In contrast to bandgap solid-state materials, it is even possible for multiple photons to be simultaneously localized by a single emitter~\citep{Rupasov_PRL1996, Sajeev_PRL1997, Calajo_PhysRev2016, Tao_PRX2016}, the formation of multi-excitation bound states lying at the heart of many interacting quantum systems~\cite{Kraemer_Nature2006, Coldea_Science2010, Fukuhara_Nature2013}.
Coupling an array of atoms through their photonic bound states offers a promising platform for investigating quantum spin models with tunable long-range interactions~\cite{Douglas_NatPhot2015, Bello_PRXQ2022}---an appealing avenue already explored in various platforms: cold atoms embedded in nanophotonic waveguides~\cite{Hood_PNAS2016, Chang_RMP2018} or optical cavities~\cite{PeriwalNature2021}, matter-wave emitters in optical lattices~\cite{KrinnerNature18}, and superconducting circuits~\cite{Sundaresan_PRX2019, Kim_PRX2021, Scigliuzzo_2021, Zhang_2022}.

The task of building a \textit{many}-body quantum optical platform has reached a complexity on par with quantum computers: you need to independently control and read out \textit{many} individual atoms (qubits). To provide a compelling architecture where the hardware overhead for control and measurement need not scale with the system size, we shift our focus towards a new paradigm that relies on coupling only a \textit{single} atom to the \textit{many} harmonic oscillator degrees of freedom in a photonic crystal. These photonic structures have a tailored dispersion relation to form optical bandgaps, where the in-band harmonic modes are energetically addressable and bestowed with strong photon-photon interactions inherited from the atom nonlinearity. The atom also needs to facilitate the preparation and characterization of quantum many-body states of light. It thus becomes important to engineer light-matter interactions where many-body effects dominate over the single-particle description.

Circuit quantum electrodynamics (cQED) is an ideal platform for pushing the light-matter coupling to novel frontiers in quantum optics~\cite{Devoret_AnnPhys2007, Diaz_RMP2019, Kockum_NatRev2019}. When the coupling strength becomes a sizable fraction of the excitation energies of the noninteracting system, the rotating wave approximation (RWA) breaks down and the resurgence of counter-rotating terms gives rise to fascinating phenomena, such as the formation of an entangled atom-photon ground state~\cite{Ciuti_PRB2005}. This regime, termed the ultrastrong coupling regime, has been demonstrated with a superconducting flux qubit coupled to a single-mode microwave resonator~\cite{Niemczyk_NatPhys2010, Diaz_PRL2010}, and has been pushed to the nonperturbative regime that captures the quantum Rabi model~\cite{Yoshihara_NatPhys2017}.
Extending this coupling regime to many harmonic modes provides an exciting avenue for investigating quantum impurity models in condensed matter physics~\cite{LeHur_PRB2012, Goldstein_PRL2103, Peropadre_PRL2013, Gheeraert_PRA2018}. Qubits, acting as quantum impurities, have been connected to high-impedance transmission lines with a discrete spectrum of modes~\cite{Puertas_npj2019, Kuzmin_npj2019, Leger_NatComm2019, Kuzmin_PRL2021, Mehta_2022} and to the electromagnetic continuum of a waveguide~\cite{Diaz_NatPhys2017, Magazzu_NatComm2018}.
These platforms thus far only probe the physics of spin-boson and Luttinger liquid models, where the engineered bosonic environment has a gapless linear dispersion. 
Realizing the ultrastrong light-matter interaction in a structured multimode vacuum, with a photonic bandgap, opens new avenues for exploring unique phenomena and capabilities beyond the reach of linear waveguide models, such as accessing nonlinear quantum optics with a single photon~\cite{Sanchez_PRL2014}, triggering single photons from an entangled vacuum~\cite{Sanchez_PRL2019}, and enabling perfect state transfer between distant atoms~\cite{Sanchez_PRA2020}.

In this work, we achieve the ultrastrong light-matter coupling regime in a photonic crystal waveguide where we explore the interplay of band-edge physics with many-body effects arising from interactions that break particle number conservation.
Our platform employs a highly nonlinear artificial atom, the fluxonium circuit, with engineered control over its internal energy levels and its coupling to the photonic crystal. We implement this model with superconducting circuits and reach the ultrastrong coupling regime using a galvanic connection.
As predicted~\cite{Sanchez_PRL2014}, the breakdown of the RWA leads to the strong hybridization of different excitation sectors which converts the transport of a single photon into a many-body problem and allows the direct observation of multi-particle bound states with a single-excitation probe.
This platform also harnesses the discrete multimode structure of the photonic crystal to microscopically probe the photonic modes and their underlying correlations using well-established quantum optics techniques. We probe the photon-photon interactions mediated by the artificial atom through a multimode fluorescence measurement and observe the broadband emission of entangled pairs of photons.

In Section~\ref{sec:model} we introduce the circuit platform and how it maps to the photonic crystal model. In Section~\ref{sec:scattering} we probe the single-photon scattering dynamics to characterize the system and the underlying multi-photon bound states. Finally, in Section~\ref{sec:MMentanglement} we investigate the multi-photon response through fluorescence measurements and probe two-mode entanglement from field correlations. 

\section{Photonic crystal platform}\label{sec:model}

Our quantum optical platform depicted in \Figref{fig:1}a consists of a multi-level artificial atom coupled to a photonic crystal waveguide with a finite-bandwidth dispersion. The discrete waveguide is represented as an array of coupled cavities, where the impurity is directly dipole-coupled to one cavity site. The Hamiltonian is given by

\begin{align*} \label{eq:H-tb}
\mathbf{H}/\hbar  = &\sum^{M-1}_{j=0} \omega_j a^\dagger_j a_j - \sum_{\langle i,j\rangle}  J_{ij}\left(a_i^\dagger a_j + a_j^\dagger a_i\right)\\
+ &\sum_l \varepsilon_{l} |l\rangle\langle l| + \sum_{l,l^\prime}  g_{ll^\prime} \sigma_{ll^\prime} \left(a_0^\dagger + a_0\right).
\end{align*}

The first two terms represent the tight-binding model for the photonic crystal, where $a^\dagger_j$ is the bosonic creation operator for a photon in the cavity site $j$. Each cavity has a bare resonance frequency $\omega_j$ and is coupled to its nearest neighbors with a tunneling rate $J_{ij}$. The third term describes the energy spectrum $\varepsilon_l$ of the uncoupled atom in its eigenbasis $|l\rangle$. The last term in the Hamiltonian describes the coupling between the atomic dipole transitions $\sigma_{ll^\prime} = |l\rangle\langle l^\prime| + |l^\prime\rangle\langle l|$ and the local cavity field in the site $j = 0$. For the range of coupling strengths $g_{01}$ investigated in this work, the counterrotating contributions to the dipole coupling are purposefully retained.

Adopting the cQED platform, we have implemented the device shown in \Figref{fig:1}c, with a circuit diagram (\Figref{fig:1}b) that directly maps to our physical model (see Appendix \ref{Sec:circmodel}).
The photonic crystal consists of a linear chain of $M=26$ lumped-element microwave resonators with bare frequencies $\omega_j/2\pi \simeq 6.9\,\text{GHz}$ and nearest-neighbor capacitive couplings $J_{ij}/2\pi \simeq 814.4\,\text{MHz}$.
We employ the fluxonium circuit~\cite{Manucharyan_Science2009} as our highly nonlinear quantum impurity. 
The qubit consists of a Josephson junction with energy $E_J/h = 8.17\,\text{GHz}$, shunted by a capacitor and an inductor, defined by a charging and inductive energy $E_C/h = 3.30\,\text{GHz}$ and $E_L/h = 5.55\,\text{GHz}$, respectively.
Given its large anharmonicity and non-trivial selection rules, this circuit becomes an ideal choice for realizing a multi-level artificial atom whose spectrum and dipole matrix elements can be controlled in situ using an external magnetic flux $\Phi_{\text{B}}$ threading the fluxonium loop.

This choice of superconducting qubit and coupling topology is favorable for reaching the ultrastrong coupling regime~\cite{Devoret_AnnPhys2007, Manucharyan_JPhysA2017, Diaz_RMP2019}.
The fluxonium circuit is galvanically coupled to a lattice site by sharing a portion of its shunt-inductor, thereby coupling the resonator current to the phase drop $\op{\varphi}_q$ across the Josephson junction. 
This becomes analogous to a magnetic dipole interaction of the atom with the resonator magnetic field ${d} {B}$, with the dipole moment defined in the fluxonium eigenbasis $d_{ij} = \langle i|\op{\varphi}_q|j\rangle$. The normalized inductive coupling strength takes the general form $g_{ij}/\omega_0 \propto \beta_L \sqrt{Z_{\text{vac}}/Z_r}\alpha^{-1/2} d_{ij}$ (see Appendix \ref{sec:galvanic_coupling}), where $\beta_L$ is the relative inductive participation ratio, $Z_{\text{vac}} = \sqrt{\mu_0/\varepsilon_0} \simeq 377\,\textOmega$ is the vacuum impedance, $Z_r$ is the resonator impedance, and $\alpha \simeq 1/137$ is the fine structure constant. To achieve large coupling strengths, we operate the fluxonium near the sweet spot $\Phi_{\text{B}} = \Phi_0/2$, where $\Phi_0=h/2e$ is the magnetic flux quantum. At this bias point, the dipole moment for the ground to first excited state transition reaches its maximum value.
Remarkably, the inductive coupling scales inversely with the fine structure constant~\cite{Devoret_AnnPhys2007}, allowing us to effortlessly reach the ultrastrong coupling regime. This is in contrast with the capacitive coupling of a superconducting qubit or coupling of a Rydberg atom to a cavity, where the small fine structure constant fundamentally limits the interaction strength.
Furthermore, we control the qubit-metamaterial coupling by implementing the shared inductor as a chain of five superconducting quantum interference device (SQuID) loops threaded by a flux $\Phi_{\text{C}}$. The SQuIDs are operated in the linear regime and jointly act as a flux-tunable inductor.

\section{Nonperturbative transport of a single photon}\label{sec:scattering}

\begin{figure}[!t]
	\begin{center}
        \includegraphics[width=0.5\textwidth]{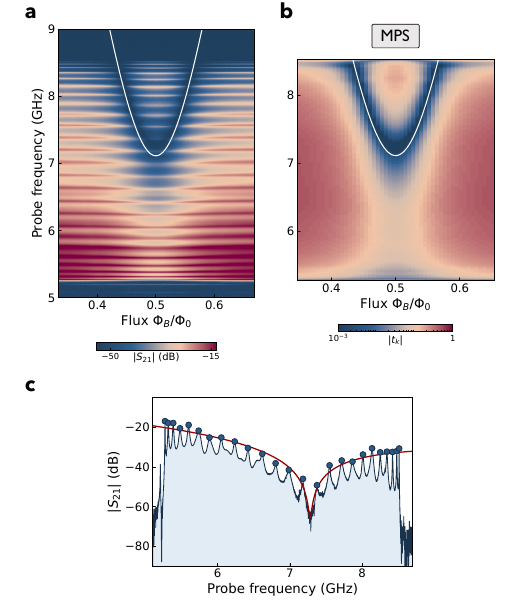}
	\end{center}
    \vspace{-0.5cm}
    \caption{\label{fig:2}\textbf{Spectroscopy near the impurity resonance.} 
    \textbf{a.}~The coupled fluxonium-metamaterial system is characterized by measuring the magnitude of microwave transmission as a function of probe frequency. The fluxonium transition is tuned in resonance with the lattice passband by varying the external flux ($\Phi_B$) through the qubit loop. The elastic scattering mediated by the qubit leads to a decrease in transmission. \textbf{b.} MPS simulations reveal that the transmission minimum is blue-shifted from the uncoupled qubit frequency (solid curve), and are thus used for extracting the fluxonium circuit parameters. \textbf{c.}~The transmission trace at the flux sweet spot $\Phi_B = \Phi_0/2$ highlights the spontaneous emission rate of the qubit into the waveguide. The extracted amplitudes for each resonance peak (blue circles) are fitted to the transmission coefficient $|T|^2$ (solid red curve) for a single photon propagating in a 1D waveguide, scattered by a two-level emitter.
    } 
\end{figure}

\begin{figure*}[!t]
	\begin{center}
		\includegraphics[width=1.0\textwidth]{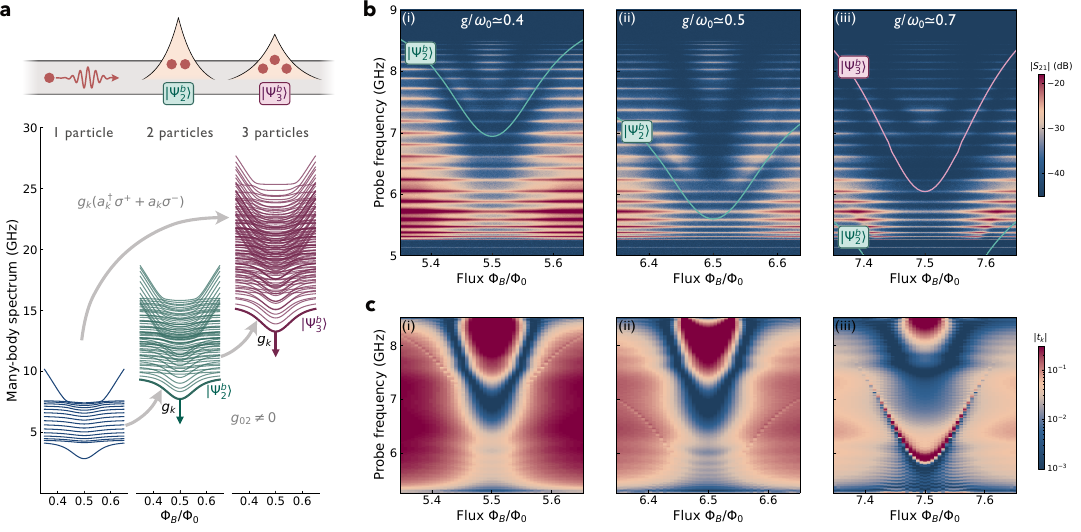}
	\end{center}
	\vspace{-0.5cm}
	\caption{\label{fig:3}\textbf{Many-body scattering dynamics of a single propagating photon.} \textbf{a.}~If the artificial atom is resticted to a two-level system and the RWA approximation is valid, the multi-particle eigenstates of the joint system form isolated bands of fixed-excitation manifolds.
	In the full model, these particle sectors are coupled through counter-rotating terms and parity-breaking transitions to higher fluxonium states, as highlighted by arrows. 
	With increased light-matter coupling, the multiphoton bound states shift down in energy, entering the single-particle band, and resonantly scatter a single propagating photon, see upper diagram where the photons are represented as red dots. \textbf{b.}~This nonperturbative effect is captured in the measured transmission amplitude of a weak probe tone for increasing values of the SQuID coupler inductance (normalized coupling strength at the flux sweet spot is listed at the top of each plot). 
	In addition to the main fluxonium transition resonance, we observe the two-photon ($|\Psi^b_{2}\rangle$) and three-photon ($|\Psi^b_{3}\rangle$) bound states entering the passband.
	The solid theory curves are the calculated bound state energy levels from diagonalizing the full model Hamiltonian. 
 \textbf{c.}~The appearance of these additional multi-particle resonances is validated with MPS scattering simulations of the transmission amplitude $t_k$. The faint resonance that gets shifted down in frequency as we increase the coupling strength qualitatively follows the calculated energy for the two-photon bound state (this resonance is present in (iii) at the lower frequency region), while the strong Fano-like resonance in (iii) matches the calculated energy for the three-photon bound state. 
	} 
\end{figure*}

The transport measurements are performed by applying a weak probe field and recording the transmitted field through the photonic crystal.
We probe the single-photon response in a heterodyne measurement using a coherent state input field with an average photon number significantly below one.
The probe frequency is varied over a broad range that covers the full dispersive band of the lattice. The transmission spectrum, shown in \Figref{fig:1}d, reveals a set of frequency-resolved resonances that correspond to the lattice modes hybridized with the fluxonium internal states. As the external magnetic bias is varied, we observe periodic disturbances in the resonance spectrum with two distinct modulation periods. The short and long periods coincide with the insertion of an additional flux quantum through the large fluxonium loop and the smaller SQuID loops, respectively. This large asymmetry $(\Phi_{\text{B}}/\Phi_{\text{C}}\approx 16)$ allows nearly independent control over the fluxonium spectrum and its coupling strength with the metamaterial. The transmission measurements are focused at flux bias values $\Phi_{\text{B}}\approx (2n+1)\Phi_0/2$ where the fluxonium resonance, corresponding to the transition from the ground state to the first excited state, enters the lattice band. Going to larger bias values, by increasing $n$, we are accessing increasing values of the coupling strength $g_{01}/\omega_0$.
In \Figref{fig:2}a we show the transmission spectrum centered around $\Phi_{\text{B}}\approx \Phi_0/2$, where the coupler SQuID is biased around its lowest inductance value. Tuning the qubit frequency $\varepsilon_{01} = \varepsilon_1 - \varepsilon_0$ across the photonic crystal spectrum leads to a clear extinction in the transmission amplitude for modes in the vicinity of the fluxonium resonance. This is explained by the destructive interference between the incoming probe field and the forward propagating field radiated by the qubit~\cite{Astafiev_Science2010}.

While in the RWA regime the transmission minimum occurs at the uncoupled qubit frequency $\varepsilon_{01}$, in the USC regime the dressed qubit frequency is renormalized by the collective coupling to the photonic modes of the waveguide. In the case of a linear waveguide, the transmission minimum is red-shifted, reminiscent of the frequency renormalization in the spin-boson model~\cite{Diaz_NatPhys2017}. In the case of our waveguide with band-edges, for the coupling regime reached at $\Phi_{\text{B}}\approx \Phi_0/2$, the transmission minimum is blue-shifted~\cite{Sanchez_PRL2014}. We validate this shift by comparing the bare qubit frequency with simulations of the scattering dynamics (see Appendix \ref{sec:MPS}) using matrix-product-states (MPS). The circuit parameters for the fluxonium circuit are thus extracted by matching the transmission data in \Figref{fig:2}a with the MPS simulations in \Figref{fig:2}b.

The coupling between the impurity and its environment is characterized by the spontaneous emission rate $\Gamma_1$ of the artificial atom into the photonic crystal waveguide. 
To quantify this rate of energy exchange, we fit the transmission spectrum at the flux sweet spot (\Figref{fig:2}c) to the model of a two-level system coupled to a one-dimensional waveguide. In the weak-driving limit, the transmission coefficient is given by $T(\omega) = 1 - \frac{1}{2}\Gamma_1/(\Gamma_2 + i\Delta)$~\cite{Astafiev_Science2010, Peropadre_NJP2013}, where $\Gamma_2$ is the decoherence rate, and $\Delta = \omega - \varepsilon_{01}$ is the detuning of the probe tone. From this simple model, we estimate a spontaneous emission rate $\Gamma_1/2\pi\simeq 6.18\,\text{GHz}$. 
This places our platform in the many-body regime, a combination of multi-mode and ultrastrong regimes, where the impurity is simultaneously exchanging excitations with many environmental modes ($\Gamma_1>\delta\omega_k$~\cite{Meiser_PRA2006, Kuzmin_npj2019}), and the interaction processes do not conserve the total number of particles ($\Gamma_1/\omega_{01} \simeq 0.85$~\cite{Diaz_NatPhys2017}). We characterize this system in terms of $\Gamma_1$ since it is readily available from the transmission spectrum, but similar conditions can be written for the coupling strength $g_{01}$~\cite{Peropadre_PRL2013}.
Thus far we have quantified this regime of light-matter interaction using a linear waveguide model~\cite{Meiser_PRA2006, Goldstein_PRL2103, Gheeraert_PRA2018}, where the fit accuracy in \Figref{fig:2}c is limited by the discrete mode structure. Nevertheless, the central observations presented in the remainder of this work offer a different physical picture that highlights the ultrastrong coupling physics and correlated nature in our structured multimode system. 

In the presence of band-edges in the photonic crystal dispersion, the ultrastrong coupling regime introduces nonperturbative modifications to the propagation of a single photon.
To understand the transmission of a single photon in our setting, we begin the discussion from the simple limit, not relevant to our experiment, where only the two lowest energy levels ($\ket{0},\ket{1}$) of the fluxonium are accounted and we neglect the counter-rotating terms in \cref{eq:H-tb}. In that case, the Hamiltonian conserves the total excitation number $N = \sum_j a_j^\dagger a_j+\ket{1}\bra{1}$, and the full eigenstate spectrum (see~\Figref{fig:3}a) consists of bands of scattering and bound states, for each integer $N$. The scattering of a single incoming photon is therefore confined to the single-excitation sector.  The higher excitation eigenstates do not participate in the dynamics, even if they are energetically accessible and lie within the single-particle band. The situation changes when the light-matter interaction is increased so that the counter-rotating terms in \cref{eq:H-tb} cannot be neglected. The Hamiltonian no longer conserves the total excitation number, but it conserves the parity $e^{i\pi N}$, meaning that the counter-rotating terms only couple sectors of the same parity (see \cref{fig:3}a). This remaining parity symmetry is broken due to the multi-level structure of the fluxonium, and the small coupling between states $\ket{0}$ and $\ket{2}$ from nontrivial selection rules.

The counter-rotating terms and the parity-breaking couplings have a striking effect on the scattering dynamics, particularly when the higher excitation bound states appear inside the single-particle band. These localized bound states, illustrated in \Figref{fig:3}a, hybridize with the single-particle scattering states, leading to Fano-like resonances in the scattering spectra~\cite{Sanchez_PRL2014}.
We directly observe this effect by probing the transmission through the lattice at larger coupling strengths achieved by increasing the inductance of the shared SQuID coupler.
The flux-dependent values for the coupling inductor, used for estimating the normalized coupling strength $g_{01}/\omega_0$, are inferred from the red-shift in the lowest lattice eigenmode (see Appendix \ref{sec:galvanic_coupling}).
In \cref{fig:3}b(i),(ii), we observe these additional resonances coming from the two-particle bound state, while for even stronger coupling the three-particle bound state also appears in the band, seen in \cref{fig:3}b(iii). 
While the counter-rotating terms are also present in \cref{fig:2}a, we only start to observe the multi-particle bound states in \cref{fig:3}b owing to the stronger inductive coupling that pushes these states to lower energy manifolds until they overlap with the single-particle band.
The observed resonances are in qualitative agreement with the calculated spectrum (theory curves in \cref{fig:3}b) for the full model using the Hamiltonian parameters of the circuit. The diminished transmission amplitude around the bound state frequencies reveals how these multi-particle states strongly scatter a single photon.
Note that the resonance in \cref{fig:3}b(iii) coming from the three-particle bound state is much more pronounced than the two-particle equivalent. This arises from the fact that the coupling between the three-particle bound state and the single-particle scattering states comes from the strong counter-rotating terms, whereas the two-particle bound state resonances are the result of the weak parity-breaking terms.

We consolidate these transmission measurements with simulations of the scattering dynamics using the matrix product state representation of the many-body wavefunction. The simulations capture the observed reflections at the bound state frequencies, as shown in Fig.~\ref{fig:3}c. Additionally, we compare these numerical results with simulations of the simplified RWA model (see Appendix \ref{sec:MPS}) and confirm that the bound state resonances are visible in the transmission spectrum only when counter-rotating terms are included in the Hamiltonian.
It becomes clear from our experimental observations and numerical results that the RWA waveguide QED model, where single photons are only reflected when resonant with the atom~\cite{Shen_PRL2005}, fails to explain the pronounced bound-state scattering, providing the signature feature of ultrastrong coupling physics in our platform.


\section{Stimulated emission of entangled photons}\label{sec:MMentanglement}

\begin{figure*}[!t]
	\begin{center}
		\includegraphics[width=1.0\textwidth]{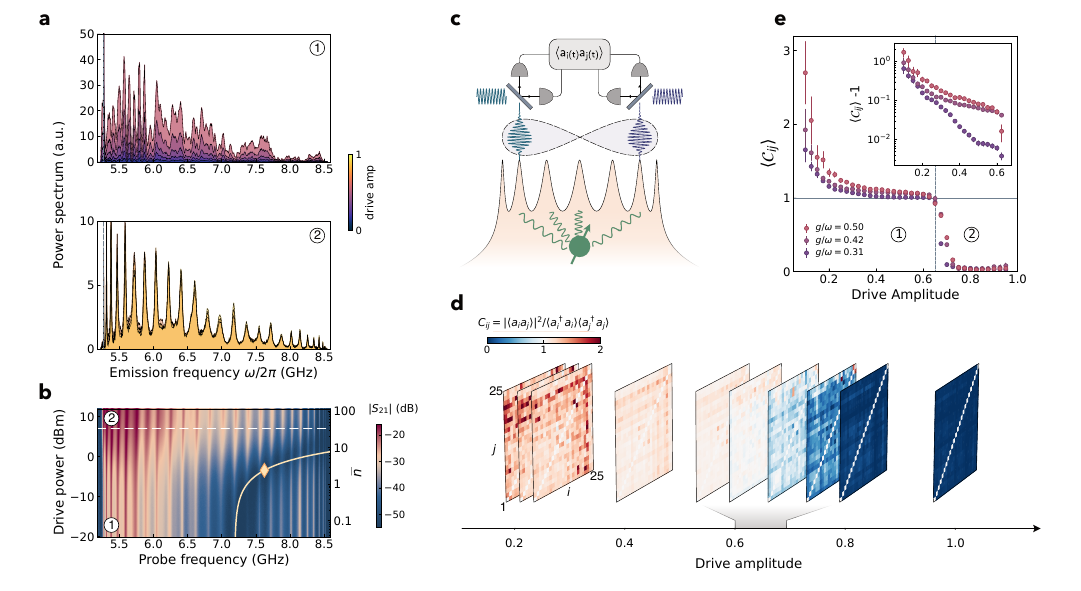}
	\end{center}
	\vspace{-0.5cm}
	\caption{\label{fig:4}\textbf{Fluorescence of multimode correlated states.} The coupling with the highly nonlinear fluxonium circuit leads to effective interactions between the modes of the photonic crystal. These interactions are reflected in the multimode fluorescence shown in \textbf{a}. Driving the lowest lattice mode (vertical line) leads to inelastic emission over the entire passband, inferred from the measured power spectrum. 
	Two emission regimes are present over the range of applied drive amplitudes: \circlenum{1} in the weak drive limit, the fluorescence spectrum displays broad peaks that closely overlap with the dressed lattice modes, \circlenum{2} in the strong drive limit, the emission profile decreases in amplitude and has sharp peaks at the unperturbed mode frequencies. \textbf{b.}~A two-tone transmission measurement reveals that the transition to the second emission regime (horizontal dashed line) coincides with the qubit resonance being Stark shifted out of the passband (solid theory line) from populating the driven lattice mode with an average of $\bar{n}$ photons.
	The marker (diamond) highlights the lowest drive amplitude used in the experiment to observe the fluorescence signal above the noise floor of the amplifier chain.
	\textbf{c.}~The quantum correlations between the emitted microwave photons are quantified using the extracted second order moments in a dual-heterodyne measurement. \textbf{d.}~We determine if any pair of modes are entangled by measuring their squeezing correlations $C_{ij}$ (for $i\neq j$) as a function of drive amplitude. \textbf{e.}~Averaging these correlations across the multimode spectrum reveals that pairs of entangled photons are being emitted in the weak drive regime \circlenum{1}, as justified by the Hillery-Zubairy criterion $\langle C_{ij}\rangle > 1$. Error bars represent one standard deviation; where absent, they are smaller than the data points.
	} 
\end{figure*}

Thus far, we have probed the elastic single-photon transport and shown it becomes a many-body problem owing to interactions that break the RWA. We expect the transport of multiple photons to become an even more complex many-body problem, a problem which we tackle in the remainder of this work by probing the inelastic response of the system, in order to characterize the qubit-mediated multimode interactions.
Early work reaching the strong coupling regime in a multimode cavity has observed the phenomena of multimode fluorescence~\cite{Sundaresan_2015}, where strongly driving the dressed qubit leads to emission over many modes close in frequency to the Rabi sidebands. In our platform, owing to the enhanced light-matter coupling, we observe complex multimode emission over the entire lattice passband, at much lower drive powers.
In \Figref{fig:4}a, as we drive the lowest lattice mode, the measured power spectrum reveals two distinct emission regimes for varying drive amplitudes (in normalized DAC units). For low drive strengths, there is a broad emission profile with a subset of the fluorescence peaks close in frequency to the dressed lattice modes. At higher drive strengths there is a dramatic decrease in the emission signal, and the fluorescence spectrum reveals sharp peaks matching the unperturbed modes shown in~\Figref{fig:1}d.

The transition between these two emission regimes is investigated with a two-tone spectroscopic measurement, where we monitor the transmission over the lattice band with a weak probe and vary the power of the additional fixed-frequency drive used in the fluorescence experiment. 
In \cref{fig:4}b, we observe that above the threshold drive strength, which separates the two emission regimes, the qubit experiences a large frequency shift that pushes its resonance outside the passband of the photonic crystal. 
This corresponds to an ac-Stark shift induced by populating the driven mode with a coherent state having an average number of $\bar{n}$ photons (see Appendix \ref{sec:stark_calib}), and we use this measurement to provide an estimate for $\bar{n}$. We find that in the weak drive regime, to observe fluorescence above the noise threshold we need to insert $\bar{n}\ge 3$ photons, and we transition into the strong drive regime with $\bar{n} > 40$ photons.

Given the large system size and the number of inserted photons, numerically reproducing the steady state nonlinear response in \cref{fig:4}a is a fascinating but challenging task that we leave for future work.
Nevertheless, we can obtain some theoretical understanding in the semi-classical limit, by modeling the measurements via input-output theory~\cite{clerkRMP_2010} and describing the system with a mean-field ansatz (see Appendix \ref{sec:MM_meanfield}). We analyze the output field of the driven system by solving the equations of motion for the full USC and the simplified RWA models, and find that the counter-rotating interaction terms and fluxonium multi-level structure are necessary precursors for observing the nonlinear multimode dynamics.
These stimulated multiphoton processes, in which photons injected in the lowest lattice mode are being converted to photons in other modes, arise from the qubit-induced interactions between the modes. Detuning the qubit transition away from the band leads to a diminished photon nonlinearity and thus a dramatic decrease in the fluorescence signal.
The simultaneous fluorescence of all modes for a fixed drive power is intriguing in itself, as it suggests the emergence of multimode correlations.

Quantifying the entanglement in such a large system is challenging, measuring the full density-matrix becomes experimentally impractical, and we need to adopt methods accessible with realistic resources.
We characterize the quantum state of the multimode output and the underlying correlations by analyzing the moments of the propagating microwave fields~\cite{Eichler_PRL2011}, where the observables are quadratic in the mode creation and annihilation operators.  
The amplified output signal is split into two separate heterodyne detection setups, with separate local oscillators selectively tuned in frequency to simultaneously measure the conjugate quadrature components for every pair of propagating modes emitted from the lattice (see \Figref{fig:4}c). This allows us to extract the second order moments of the complex field amplitudes and evaluate the field correlations by subtracting the amplifier noise moments (see Appendix \ref{sec:MM_HZ}).  
We examine the entanglement structure by employing the Hillery-Zubairy criterion~\cite{Hillery_Zubairy_PRL2006}, which states that any (pure or mixed) two-mode separable state, defined by the bosonic annihilation operators $a$ and $b$, satisfies the condition $|\langle a b\rangle|^2 \le \langle a^\dagger a\rangle \langle b^\dagger b\rangle$. 
When the quantum correlations on the left hand side become larger than the product of intensities, the state is identified as entangled, favoring a superposition of states that differ by a photon in each mode similar to a two-mode squeezed state. This is a sufficient but not necessary condition for entanglement, as there are other classes of two-mode entangled states that satisfy the above separability inequality. This entanglement criterion is tested on our multimode state by measuring the squeezing correlations $C_{ij} = |\langle a_i a_j\rangle|^2/\langle a_i^\dagger a_i\rangle \langle a_j^\dagger a_j\rangle$ for the fields emitted at every lattice mode frequency. In~\Figref{fig:4}d, we highlight the evolution of the full correlation matrix as a function of the pump tone amplitude (see Appendix \ref{sec:MM_HZ}). Our global measure of entanglement $\langle C_{ij} \rangle$ is extracted from averaging all $i\neq j$ mode-pair correlations for each matrix. The dependence of this metric on the pump amplitude is displayed in~\Figref{fig:4}e for three distinct values of the coupling strength. We find that, in the low drive strength regime, the two-mode correlations are on average above one, which violates the separability criterion. This reveals how the nonlinear wave-mixing processes stimulated by the pump tone lead to inelastic emission of entangled pairs of photons.


\section{Outlook}

In this work, we have demonstrated the ultrastrong coupling of a highly nonlinear emitter to a photonic crystal waveguide, where multi-photon bound states modify the transmission of photons in the lattice and provide a new avenue for exploring nonlinear quantum optics at the single-photon level~\citep{Sanchez_PRL2014, Kockum_PRA2017}. 
Similar to the single-mode picture, the ground state contains a multimode cloud of virtual photons centered around the emitter~\cite{Peropadre_PRL2013}. Harnessing the full control over the emitter's coupling to its environment, these vacuum fluctuations can be converted into single-photon radiation by modulating the coupling strength~\cite{Sanchez_PRL2019}, a process which is again mediated by the photon bound states. 
This platform can also be extended to the frustrated ultrastrong coupling of an impurity to two competing baths~\cite{Belyansky_PRR2021}, where the dressed-spin quasiparticle description breaks down and the induced photon-photon interactions can be highly anisotropic.
Exploiting the three-wave mixing nonlinearity of the fluxonium circuit, this multimode platform can also be employed as a quantum simulator in synthetic dimensions, where frequency-selective drives can induce inter-mode particle hopping and blockade-induced interactions~\cite{Vrajitoarea_NatPhys2020, chakram2022multimode}.

Finally, the multimode correlation measurements presented in this work can become a useful technique for probing entanglement in large-scale quantum systems, further expanding the quantum optics toolbox for characterizing strongly-correlated photonic materials~\cite{Carusotto_NatPhys2020}.
The sharp change in the squeezing correlations observed in~\Figref{fig:4} could in fact be indicative of a driven-dissipative phase transition between the two inelastic emission regimes~\cite{Fitzpatrick_PRX2017}.
Furthermore, this pumped nonlinear system can be used as a broadband correlated reservoir for quantum communication applications, where the correlated photon fields can drive distant qubit nodes and distribute entanglement in a quantum network~\cite{KrausCirac_PRL2004, Agusti_arxiv2022}. This stimulated reservoir can also be potentially used as a source of multimode squeezed light in 
boson sampling experiments for demonstrating quantum computational advantage~\cite{Deshpande_SciAdv2022}. 
\\
\begin{acknowledgments}
This work was supported by the NSF (PHY-1607160) and the MURI (W911NF-15-1-0397). R.B., R.L., S.W., and A.V.G.~acknowledge support by NSF QLCI (award No.~OMA-2120757), DoE QSA, DARPA SAVaNT ADVENT, ARO MURI, the DoE ASCR Quantum Testbed Pathfinder program (award No.~DE-SC0019040), DoE ASCR Accelerated Research in Quantum Computing program (award No.~DE-SC0020312), NSF PFCQC program, AFOSR, AFOSR MURI. Devices were fabricated in the Princeton University Quantum Device Nanofabrication Laboratory (QDNL) and in the Princeton Institute for the Science and Technology of Materials (PRISM) cleanroom. The authors acknowledge the use of Princeton’s Imaging and Analysis Center, which is partially supported by the Princeton Center for Complex Materials, a National Science Foundation (NSF)-MRSEC program (DMR-1420541).
\end{acknowledgments}





\appendix

\section{Experimental setup}\label{Sec:expsetup}
\subsection{Device Design and Fabrication}

\begin{figure*}[!t]
	\begin{center}
		\includegraphics[width=1\textwidth]{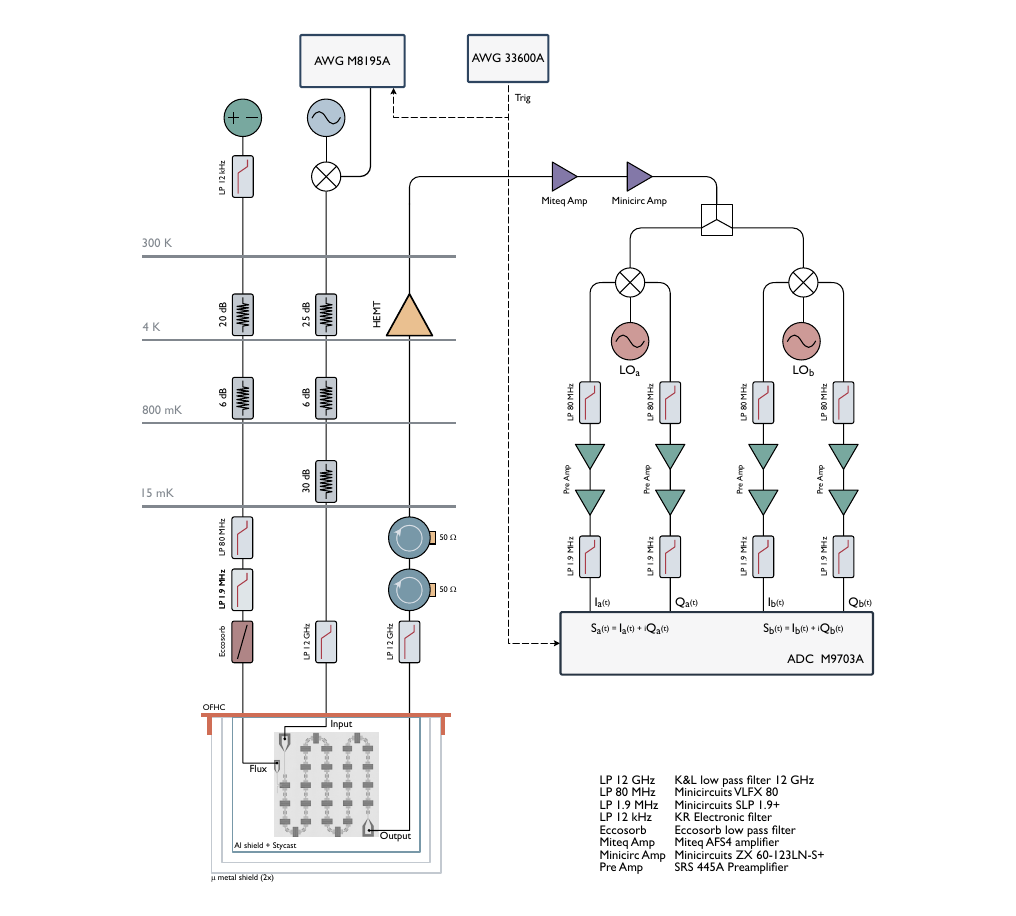}
	\end{center}
	\vspace{-0.5cm}
	\caption{\label{fig:ExpSetup}\textbf{Experimental setup}. Schematic diagram of the cryogenic and instrumentation setup.} 
\end{figure*}

The photonic crystal is experimentally implemented using 26 lumped-element microwave resonators, capacitively connected in a chain as shown in \Figref{fig:1}c. The lumped element resonator geometry was optimized using commercial software. The capacitive elements are implemented as interdigitated structures where the values for both the ground and coupling capacitances are calibrated using ANSYS Maxwell (electrostatic solver). The resonator inductors are implemented as meandered microstrip wires with a 4 {\textmu}m width. The inductance values are inferred from calibrating the resonance frequency of the resonator site using ANSYS HFSS (eigenmode solver) and AWR Microwave Office (AXIEM solver). The inductance of the resonator site coupled to the fluxonium is intentionally made smaller to accommodate the inductive contribution from the fluxonium coupling inductor. The photonic crystal circuit is fabricated using a 200~nm-thick Nb film sputtered on a 525~{\textmu}m-thick high-purity C-plane sapphire substrate. The components are defined using optical lithography and reactive ion etching, with an attainable minimum feature size of 1{\textmu}m.  The wafer was diced into individual 7x7 $\text{mm}^2$ chips.

The fluxonium artificial atom is composed of a small Josephson junction ($100\,\text{nm} \times 220\,\text{nm}$) inductively shunted by an array of 38 larger Josephson junctions ($190\,\text{nm} \times 1500\,\text{nm}$) and a tunable inductor, implemented as a chain of four asymmetric SQuIDs. The junctions and superconducting loop were defined via electron-beam lithography using a bi-layer resist MMA(methyl methacrylate)-PMMA(poly methyl methacrylate). The junctions were fabricated via double-angle electron-beam evaporation using a Dolan bridge technique. Aluminum films, with a thickness of 30~nm and 60~nm respectively, were deposited at different angles. The junction electrodes were separated by an AlO\textsubscript{x} oxide grown at ambient temperature for 10 minutes in 200~mbar static pressure of an Ar:O\textsubscript{2} (85\,\%:15\,\%) gas mixture. After the evaporation of the second aluminum layer, a final oxide was grown using the same gas mixture for 20 minutes at 40~mbar, in order to passivate the surface. After evaporation, the remaining resist was lifted off by leaving the sample in N-Methyl-2-Pyrrolidone (NMP) at 80\,\degree C for 3 hours. The fluxonium is inductively coupled to the edge resonator on the input port of the device. The contact pads for the galvanic connection between the resonator inductor and the fluxonium loop are intentionally made larger ($20\times50~\text{\textmu m}^2$) than the inductor width to ensure a small contact resistance.

The fluxonium loop and tunable inductor are magnetically biased using an on-chip flux line where we apply a DC current. The magnetic field induced by the DC current is modeled using the Biot-Savart law, and integrating the field over the area of the loops gives an estimate for the applied magnetic flux. The area of the fluxonium and SQuID loops, as well as their distance with respect to the flux bias line, are designed to achieve the desired ratio of applied flux. Specifically, for a given applied current bias, we want the magnetic flux enclosing the fluxonium loop to be at least an order of magnitude larger than the magnetic flux in each SQuID loop. This large asymmetry allows the flexibility of biasing the fluxonium artificial atom to have the same energy spectrum while sampling various coupling regimes. Taking into consideration screening currents due to the Meissner effect is not critical for this analysis since we are interested in the ratio of applied flux instead of the absolute values.

The parameters of the circuit are summarized in Table~\ref{table:CircParams} using the circuit notation outlined in the main text. The parameter values are inferred from fitting the experimental data to the circuit model in section~\ref{Sec:circmodel}.

\begin{table}[!t]
\centering
\begin{tabular}[t]{lc}
\hline
Fluxonium & \\
\hline
$E_J/\hbar$& 8.17 GHz\\
$E_L/\hbar$& 5.55 GHz\\
$E_C/\hbar$& 3.30 GHz\\
\hline
Resonator chain & \\
\hline
$L_r$& 2.80 nH\\
$L_r^\prime$ (edge)& $\sim$ 0.1nH\\
$L_c$& (4 to 14) nH\\
$C_g$& 249.15 fF\\
$C_c$& 202.70 fF\\
\hline
\end{tabular}
\linespread{1.2}\caption{\label{table:CircParams}{Fluxonium and photonic crystal circuit parameters.}}
\end{table}

\subsection{Cryogenic Setup and Control Instrumentation}

The device was mounted to the base stage of a dilution refrigerator, as shown in \Figref{fig:ExpSetup}. Device shielding consists of two \textmu-metal layers and an inner aluminum layer covered with Stycast. The cryogenic setup, including attenuation and filtering, is shown schematically in \Figref{fig:ExpSetup}. The resonator chain is connected to two coax cables, an input line used for microwave driving, and an output line used for measuring the transmitted fields through the waveguide. The flux bias current is sent through a separate control line with a bandwidth of 12 kHz. The transmitted signal passes through two cryogenic isolators, thermalized at the base stage, and is amplified using a high-electron-mobility transistor (HEMT) amplifier, anchored at the 4 K stage.

The elastic transmission experiments were performed using a Network Analyzer (Keysight N5241A PNA-X) and the inelastic emission experiments were performed using the pulsed setup shown in \Figref{fig:ExpSetup}. The microwave pulses for driving the photonic crystal waveguide are generated using a vector signal generator (Keysight E8267D), with internal wide-band IQ mixing functionality. The base-band and marker pulses are generated with an arbitrary waveform generator (AWG Keysight M8195A). The flux bias current is provided by a low-noise current source (YOKOGAWA GS200).
The signal coming out of the fridge is amplified by low-noise room temperature amplifiers. The quadrature components of the output signal are measured with a standard two-channel homodyne setup, using an IQ mixer (Marki IQ4509LXP) and a local oscillator set in the frequency range of the eigenmode band. After filtering and amplifying the homodyne signal, we digitize it using an analog-to-digital converter (ADC) card (Keysight M9703A) with a 1 GS/s sampling rate and 12-bit voltage scale resolution.

For the multi-mode correlations, we split the output signal into two homodyne measurement branches, with separate local oscillators set to the desired frequencies of the eigenmode fields we want to measure correlations of. The homodyne voltages in the two branches, $\{I_a(t), Q_a(t)\}$ and $\{I_b(t), Q_b(t)\}$, are extracted simultaneously using the four channels of the ADC card, and combined to yield the complex field amplitudes $S_{a,b}(t) = I_{a,b}(t) + i Q_{a,b}(t)$. The complex fields are used for evaluating the second order moments $\langle S^\ast_a S_a\rangle, \langle S^\ast_b S_b\rangle, \langle S_a S_b\rangle$.

\section{Circuit model}\label{Sec:circmodel}
The lumped-element circuit model for the photonic crystal is described using various methods that involve identifying the waveguide eigenmodes and mapping the circuit to a tight-binding lattice model.

\subsection{Chain of coupled cavities \label{circ-cca}}
\begin{figure}[!t]
	\begin{center}
		\includegraphics[width=0.5\textwidth]{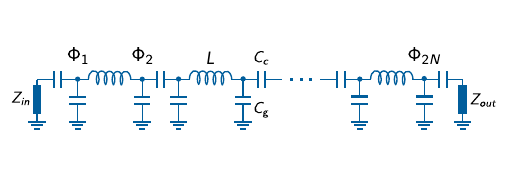}
	\end{center}
	\vspace{-0.5cm}
	\caption{\label{fig:CCA_circ}\textbf{Circuit diagram for a discretized photonic crystal}. Schematic diagram of a one-dimensional chain of capacitively coupled microwave resonators.} 
\end{figure}

Using the well-established circuit quantization formalism \citep{Devoret_LesHouches}, we describe the eigenmodes of the resonator array in terms of its Lagrangian. The circuit diagram of the bare (qubit-less) resonator chain is shown in \Figref{fig:CCA_circ}. The chain consists of $N$ lumped-element resonators, composed of inductors of inductance $L$ and capacitors of capacitance $C_g$, and their voltages are coupled through a series capacitor of capacitance $C_c$. This circuit has $2N$ degrees of freedom, equal to the number of nodes, and half of these degrees of freedom are relevant for this experiment. In the realistic experimental scenario, the resonator array is capacitively coupled to waveguides at the input and output ports, which we model as impedance terminations $Z_{\text{in}}$ and $Z_{\text{out}}$, respectively. The useful variables to describe the Lagrangian are the flux $\Phi_{n}$ and voltages $\dot{\Phi}_{n}$ at each node $n\in [1, 2N]$. The drawback of this formalism is not being able to take into account the impedance at the input and output ports, except for the limits when they tend to zero or infinity. The Lagrangian of the cavity chain for the boundary conditions $Z_{\text{in}}$, $Z_{\text{out}} \rightarrow 0$ is given by
\begin{dmath}\label{Lcca}
\mathcal{L}_{\text{cca}} = \sum_{{n} = 1}^{N} \left[ 
\frac{C_\Sigma}{2}\left( \dot{\Phi}^2_{2n-1} + \dot{\Phi}^2_{2n} \right)
- C_c \dot{\Phi}_{2n}\dot{\Phi}_{2n+1} 
- \frac{1}{2L}\left( \Phi_{2n} - \Phi_{2n-1} \right)^2
\right],
\end{dmath}
where the summation is performed over every resonator unit cell, and we define $C_\Sigma \triangleq C_g + C_c$. Since the chain contains only linear elements, the Lagrangian is quadratic in the coordinate variables and can be written in a compact matrix form
\begin{align}
\mathcal{L}_{\text{cca}} = \frac{1}{2} {\dot{\vec{\op{\Phi}}}}^{\text{t}} \hat{\op{C}} \dot{\vec{\op{\Phi}}} - 
\frac{1}{2} {\vec{\op{\Phi}}}^{\text{t}} \hat{\op{L}}^{-1} \vec{\op{\Phi}}
\end{align}
where we define the flux and voltage coordinate vectors
\begin{align} 
{\vec{\op{\Phi}}} =\, 
\setlength\arraycolsep{2pt}\def\arraystretch{0.6}
\begin{pmatrix}
\Phi_{1}\\
\Phi_{2}\\
\vdots\\
\Phi_{2N}\\
\end{pmatrix},\,\,
\dot{\vec{\op{\Phi}}} =\, 
\setlength\arraycolsep{2pt}\def\arraystretch{0.6}
\begin{pmatrix}
\dot{\Phi}_{1}\\
\dot{\Phi}_{2}\\
\vdots\\
\dot{\Phi}_{2N}\\
\end{pmatrix}
\end{align}
and used the capacitance $\hat{\op{C}}$ and inductance $\hat{\op{L}}$ matrices
\begin{align} 
\hat{\op{C}} =\, 
\setlength\arraycolsep{2pt}\def\arraystretch{0.6}
\begin{pmatrix}
C_\Sigma& & & & & & \\
 &C_\Sigma&-C_c & & & & \\
 &-C_c &C_\Sigma& & & &\\
 & &\ddots &\ddots &\ddots & &\\
 & & & &C_\Sigma &-C_c & \\
 & & & &-C_c &C_\Sigma & \\
 & & & & & &C_\Sigma \\
\end{pmatrix}
\end{align}

\begin{align}
\hat{\op{L}}^{-1} = \frac{1}{L}\,
\setlength\arraycolsep{2pt}\def\arraystretch{0.6}
\begin{pmatrix}
1 &-1 & & & & & \\
-1 &1 & & & & & \\
 & &1 &-1 & & &\\
 & &-1 &1 & & &\\
 & & &\ddots &\ddots &\ddots &\\
 & & & & &1 &-1 \\
 & & & & &-1 &1 \\
\end{pmatrix}\, .
\end{align}

If we focus on a single unit cell, we can define the resonator in terms of the voltage $\dot{\Phi}_{2n} - \dot{\Phi}_{2n-1}$ across the series ground capacitor and current $\Phi_{2n} - \Phi_{2n-1}$ through the inductor. This argument will also become clear when visiting the treatment for galvanically coupling the fluxonium to a unit cell. It thus becomes more intuitive to describe the resonator chain in a different coordinate basis $\Phi_{n}^\pm \triangleq \Phi_{2n} \pm \Phi_{2n-1}$, defined by \textit{differential} $\Phi_{n}^-$ and \textit{center of mass} (COM) $\Phi_{n}^+$ degrees of freedom for ${n}\in [0,{N}]$. The corresponding Lagrangian in this basis becomes

\begin{dmath}
\mathcal{L}_{\text{cca}}^\pm = \sum_{{n} = 1}^{N} \left[ 
\frac{C_\Sigma}{4}\left( \dot{\Phi}^{- 2}_{n} + \dot{\Phi}^{+ 2}_{n}\right)
- \frac{C_c}{4} \left( \dot{\Phi}^{+}_{n} + \dot{\Phi}^{-}_{n} \right)
\left( \dot{\Phi}^{+}_{n+1} + \dot{\Phi}^{-}_{n+1} \right)
- \frac{1}{2L} {\Phi}^{- 2}_{n}
\right],
\end{dmath}
which can also be written in a similar compact form
\begin{align}\label{Lccapm}
\mathcal{L}_{\text{cca}}^{\pm} = \frac{1}{2} {\dot{\vec{\op{\Phi}}}}_\pm^{\text{t}} \hat{\op{C}}_\pm \dot{\vec{\op{\Phi}}}_\pm -
\frac{1}{2} {\vec{\op{\Phi}}}^{\text{t}}_\pm \hat{\op{L}}_\pm^{-1} \vec{\op{\Phi}}_\pm
\end{align}
with basis vectors in the differential/COM coordinates
\begin{align} 
{\vec{\op{\Phi}}}_\pm =\, 
\setlength\arraycolsep{1pt}\def\arraystretch{0.3}
\begin{pmatrix}
\Phi_{1}^\pm\\
\Phi_{2}^\pm\\
\vdots\\
\Phi_{2N}^\pm\\
\end{pmatrix},\,\,
\dot{\vec{\op{\Phi}}}_\pm =\, 
\setlength\arraycolsep{1pt}\def\arraystretch{0.3}
\begin{pmatrix}
\dot{\Phi}_{1}^\pm\\
\dot{\Phi}_{2}^\pm\\
\vdots\\
\dot{\Phi}_{N}^\pm\\
\end{pmatrix},
\end{align}
and with different capacitance and inductance matrices
\begin{widetext}
\begin{align} 
\hat{\op{C}}_\pm =\, 
\setlength\arraycolsep{2pt}\def\arraystretch{0.7}
\begin{pmatrix}
C_\alpha& &C_\beta & -C_\beta & & & & & &  \\
&C_\alpha &C_\beta & -C_\beta & & & & & &  \\
C_\beta & C_\beta &C_\alpha & &C_\beta &C_\beta & & & &  \\
-C_\beta & -C_\beta & &C_\alpha &-C_\beta & -C_\beta & & & &  \\
& &C_\beta & -C_\beta &C_\alpha & & & & &  \\
& &C_\beta & -C_\beta & &C_\alpha & & & & \\
& & & & & &\ddots & &C_\beta &-C_\beta \\
& & & & & & & &C_\beta &-C_\beta \\
& & & & & &C_\beta &C_\beta &C_\alpha & \\
& & & & & &-C_\beta &-C_\beta & &C_\alpha \\
\end{pmatrix}, \qquad
\hat{\op{L}}^{-1}_\pm = \frac{1}{L}\,
\setlength\arraycolsep{2pt}\def\arraystretch{0.4}
\begin{pmatrix}
1 & & & & & & \\
&0 & & & & & \\
 & &1 & & & &\\
 & & &0 & & &\\
 & & & &\ddots & &\\
 & & & & &1 & \\
 & & & & & &0 \\
\end{pmatrix}\,,
\end{align}
\end{widetext}
for abbreviation we defined $C_\alpha = C_\Sigma/2$ and $C_\beta = C_c/4$.

The Euler-Lagrange equation of motion takes the form of a wave equation
\begin{align}\label{LagEq}
{\ddot{\vec{\op{\Phi}}}}_\pm + \hat{\op{C}}_\pm^{-1} \hat{\op{L}}_\pm^{-1} {{\vec{\op{\Phi}}}}_\pm = 0 ,
\end{align}
where the normal mode resonance frequencies $\omega_{\text{k}}$ of the chain are the positive square roots of the eigenvalues of the matrix $\hat{\op{C}}_\pm^{-1} \hat{\op{L}}_\pm^{-1}$. 
The matrices $\hat{\op{C}}^{-1} \hat{\op{L}}^{-1}$ and $\hat{\op{C}}_\pm^{-1} \hat{\op{L}}_\pm^{-1}$ are connected through a basis transformation $\vec{\op{\Phi}}_\pm = \mathbf{U}_\pm\vec{\op{\Phi}}$ and therefore have the same eigenfrequencies. Inspecting the eigenvalue spectrum, $N$ of these eigenvalues have zero frequency while the other $N$ modes have the desired cosine dispersion expected for a tight-binding chain. 
These zero-frequency eigenstates correspond to the static COM variables $\Phi_{n}^{+}$, while the dispersive band of states centered around the bare resonator frequency corresponds to the dynamics of the coupled differential variables $\Phi_{n}^{-}$.
This transformation $\mathbf{U}_\pm$ becomes convenient for rewriting the eigenvectors in a different basis that allows us to differentiate static degrees of freedom from the oscillator degrees of freedom.
Although, from Eq.~\ref{Lccapm}, the differential and COM variables are coupled to each other through their voltages, the effect on the tight-binding model for the differential degrees of freedom is a minimal, small renormalization of the hopping parameters. For the remainder of this theoretical treatment, we will not consider these zero-frequency eigenvalues and focus only on the differential normal modes of the chain.

\subsection{ABCD matrix calculation}

The photonic crystal circuit can also be characterized using the ABCD matrix formalism, used extensively for calculating the scattering parameters of a circuit with an arbitrary configuration, probed at any given frequency. Circuits with multiple ports can be conveniently analyzed by decomposing them into two-port modules connected in series. Any two-port circuit can be described in terms of a $2 \times 2$ matrix which relates the voltage $V_{\text{j}}$ and current $I_{\text{j}}$ between each port as

\begin{align} 
\setlength\arraycolsep{2pt}\def\arraystretch{0.6}
\begin{pmatrix}
V_{\text{1}}\\
I_{\text{1}}\\
\end{pmatrix}\, = \,
\setlength\arraycolsep{2pt}\def\arraystretch{0.6}
\begin{pmatrix}
A&B\\
C&D\\
\end{pmatrix}
\begin{pmatrix}
V_{\text{2}}\\
I_{\text{2}}\\
\end{pmatrix}.
\end{align}
The convenience lies in the fact that the ABCD matrix of several two-port networks connected in series is given by the product of the ABCD matrices of each network.

For the case of our photonic crystal circuit, the ABCD matrix of a single lumped-element resonator can be extracted from the individual matrices for the series inductor and parallel capacitor to ground
\begin{align}
\nonumber\mathbf{M}_{\mathrm{uc}}(\omega) &= \mathbf{M}_{\mathrm{C_g}} \mathbf{M}_{\mathrm{L}} \mathbf{M}_{\mathrm{C_g}}\\
&= \begin{pmatrix}
1&0\\
\mathrm{j}\omega C_g &1\\
\end{pmatrix}
\,\cdot\,
\begin{pmatrix}
1&\mathrm{j}\omega L\\
0&1\\
\end{pmatrix}
\,\cdot\,
\begin{pmatrix}
1&0\\
\mathrm{j}\omega C_g &1\\
\end{pmatrix}.
\end{align}

The unit cells are connected in series through a coupling capacitor, and the edges are capacitively coupled to a section $l$ of a coplanar waveguide with a characteristic impedance $Z_0 = 50\,\Omega$ and phase velocity $\upsilon_p$. The matrix terms for these additional components are given by
\begin{dmath}
\mathbf{M}_{\mathrm{C_c}} = 
\begin{pmatrix}
1&\-\mathrm{j}/\omega C_c\\
0&1\\
\end{pmatrix}
\end{dmath},
\begin{dmath}
\mathbf{M}_{\mathrm{cpw}} = \begin{pmatrix}
\cos\left( \omega l /\upsilon_p \right) & \mathrm{j} Z_0 \sin\left( \omega l /\upsilon_p \right)\\
\mathrm{j} \sin\left( \omega l /\upsilon_p \right)/Z_0 & \cos\left( \omega l /\upsilon_p \right)\\
\end{pmatrix}.
\end{dmath}

From the periodicity of the circuit, the ABCD matrix of the entire resonator chain becomes
\begin{align}
\mathbf{M}_{\mathrm{cca}}(\omega) = \mathbf{M}_{\mathrm{cpw}} \, \mathbf{M}_{\mathrm{C_c}} \, \left[ \mathbf{M}_{\mathrm{uc}} \mathbf{M}_{\mathrm{C_c}}\right]^{{N}} \, \mathbf{M}_{\mathrm{cpw}}.
\end{align}

Since we are interested in the transmission coefficient through the device, we can easily calculate the scattering parameter $S_{\text{21}}$ using the ABCD matrix of the chain
\begin{align}
S_{\text{21}}(\omega) = \frac{2}{A + B/Z_0 + C Z_0 + D}.
\end{align}

\begin{figure}[!t]
	\begin{center}
		\includegraphics[width=0.5\textwidth]{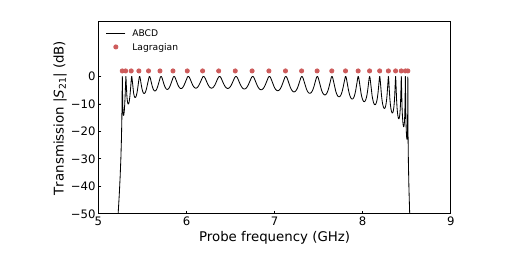}
	\end{center}
	\vspace{-0.5cm}
	\caption{\label{fig:ABCDvsLag}\textbf{Calculated transmission in a resonator chain using ABCD matrices}. The solid line (blue) corresponds to the magnitude of the transmission coefficient using the ABCD matrix approach. Data points (red) correspond to the eigenmode frequencies of the circuit Lagrangian.
.} 
\end{figure}

The advantage of this method over the Lagrangian approach is the capability of finding the eigenmodes of the circuit for arbitrary boundary conditions set by input and output port impedances $Z_{\text{in}}$ and $Z_{\text{out}}$, which are set to $50\,\Omega$ for this experiment. The eigenmode frequencies correspond to the resonances found in transmission. 
The small disadvantage of this method is that the precision of eigenmode values is set by the resolution of the frequency grid in which you are evaluating the ABCD matrices, whereas diagonalizing a $2N\times 2N$ matrix can be more efficient. We combine both methods for determining the circuit parameters of the device.

For completeness, we compare these two methods as shown in \Figref{fig:ABCDvsLag}. The eigenfrequencies of the differential modes, inferred from the eigenvalues of the $\hat{\op{C}}_\pm^{-1} \hat{\op{L}}_\pm^{-1}$ matrix, are overlayed on top of the magnitude of the transmission coefficient $|S_{21}|$ evaluated over the frequency range of the single-photon band, and are found to align with the resonance frequencies (Lorentzian peaks) in transmission. Both calculations are performed with the circuit parameters used in the experiment, displayed in Table~\ref{table:CircParams}, which yield similar eigenmode frequencies, within disorder levels.

\subsection{Tight-binding model}

The circuit is translated into a tight-binding model by moving to a Hamiltonian representation, described using the flux $\Phi_n$ and charge $Q_n$ at each node. The charge variables are conjugate momenta of the node fluxes and are found from $\vec{\op{Q}} = \partial \mathcal{L}_{\mathrm{cca}}/\partial \dot{\vec{\op{\Phi}}} = \hat{\mathbf{C}} \dot{\vec{\op{\Phi}}}$, where $\vec{\op{Q}}$ represents the basis vector for all the node charges $\left(Q_1, Q_2,\dots Q_{2N} \right)^{\text{t}}$. A similar relationship holds in the differential/COM basis  $\vec{\op{Q}}_\pm = \hat{\mathbf{C}}_{\pm} \dot{\vec{\op{\Phi}}}_{\pm}$. Using the Lagrangian in Eq.~\ref{Lccapm}, we can obtain the Hamiltonian for the resonator chain from a Legendre transformation
\begin{align}\label{Hccapm}
\mathbf{H}_{\text{cca}}^{\pm} = \frac{1}{2} \vec{\op{Q}}_\pm^{\text{t}} \hat{\op{C}}^{-1}_\pm \vec{\op{Q}}_\pm 
+ \frac{1}{2} {\vec{\op{\Phi}}}^{\text{t}}_\pm \hat{\op{L}}_\pm^{-1} \vec{\op{\Phi}}_\pm
\end{align}

Following the discussion from section~\ref{circ-cca}, we can expand the Hamiltonian in terms of differential variables, and purposefully write it in this form
\begin{dmath}
\mathbf{H}_{\text{cca}}^{\pm} = \sum_j \left( \frac{1}{2} \left[\hat{\op{C}}^{-1}_\pm\right]_{j,j} {Q^{-}_j}^2 
+ \frac{1}{2} \left[\hat{\op{L}}^{-1}_\pm\right]_{j,j} {\Phi^{-}_j}^2 \right)
+ \sum_{\langle i,j \rangle} \left[\hat{\op{C}}^{-1}_\pm\right]_{i,j} Q_i^{-} Q_j^{-},
\end{dmath}
where we decouple harmonic oscillator terms containing quadratic charge and flux variables with the same resonator coordinate, from terms that couple the charge degrees of freedom for different resonators. 

We quantize the Hamiltonian by promoting the flux $\Phi$ and charge $Q$ variables to operators that satisfy the commutation relations $\left[\mathbf{\Phi}_{n}, \mathbf{Q}_{m}\right] = \mathrm{i}\hbar\delta_{n,m}$. As we do for the case of a single harmonic oscillator, we can express the operators in terms of bosonic ladder operators
\begin{align}
\nonumber \mathbf{Q}^{-}_j = \mathrm{i}\sqrt{\frac{\hbar}{2}} \,\, Z_j^{-\frac{1}{2}} \,\, \left( \op{a}_j^\dagger - \op{a}_j \right)\\
\mathbf{\Phi}^{-}_j = \sqrt{\frac{\hbar}{2}} \,\, Z_j^{\frac{1}{2}} \,\, \left( \op{a}_j^\dagger + \op{a}_j \right),
\end{align}
where $\op{a}_j^\dagger$ ($\op{a}_j$) create (annihilate) photons in the $j^{\text{th}}$ resonator, and the resonator's characteristic impedance entering in the zero point fluctuation amplitude is given by $Z_j \triangleq \left[\hat{\op{C}}^{-1}_\pm \right]_{j,j}^{\frac{1}{2}}/ \left[\hat{\op{L}}^{-1}_\pm \right]_{j,j}^{\frac{1}{2}}$. Inserting the quantum operators into the Hamiltonian in Eq.~\ref{Hccapm} gives the following tight-binding model for describing the photonic lattice

\begin{dmath}\label{Htb}
\mathbf{H}_{\text{tb}}/\hbar = \sum_j \omega_j \left( \op{a}_j^\dagger \op{a}_j + \frac{1}{2}\right)
+ \sum_{\langle i,j \rangle} t_{i,j} \left(\op{a}_i^\dagger  - \op{a}_i \right) \left(\op{a}_j^\dagger  - \op{a}_j \right),
\end{dmath}
where the first summation accounts for the on-site energy given by the resonator frequency $\omega_j$, and the second summation describes the tunneling of microwave excitations between nearest neighbor oscillators with a tunneling rate $t_{i,j}$. These terms are extracted from the capacitance and inductance matrices in the circuit model
\begin{align}\label{termsTB}
\omega_j \triangleq \left[\hat{\op{L}}^{-1}_\pm \right]_{j,j}^{\frac{1}{2}} \left[\hat{\op{C}}^{-1}_\pm\right]_{j,j}^{\frac{1}{2}}\\
t_{i,j} \triangleq - \frac{1}{2} \left(Z_i Z_j\right)^{-\frac{1}{2}} \left[\hat{\op{C}}^{-1}_\pm\right]_{i,j}.
\end{align}
Describing the photonic crystal in this tight-binding model provides a more intuitive representation which we will adopt for modeling transport through the lattice coupled to an artificial atom.

\subsection{Galvanically coupled artificial atom}\label{sec:galvanic_coupling}

We analyze and derive the microscopic model for a fluxonium circuit, which plays the role of a highly nonlinear artificial atom, embedded in the photonic crystal. The first step is to formulate the coupling to a single unit cell and extend that to the full oscillator chain.

\subsubsection{Coupling to a single unit cell}\label{galvanicUC}

\begin{figure}[!t]
	\begin{center}
		\includegraphics[width=0.5\textwidth]{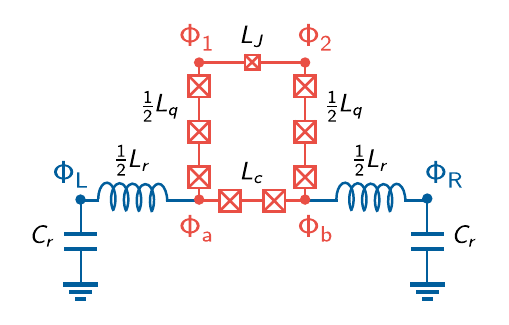}
	\end{center}
	\vspace{-0.5cm}
	\caption{\label{fig:galvanic}\textbf{Circuit diagram for a galvanically coupled fluxonium}. Schematic diagram for a fluxonium artificial atom current-coupled to a single microwave resonator, through a mutually shared inductor.} 
\end{figure}

The circuit diagram for a fluxonium qubit galvanically coupled to a single lumped-element resonator is shown in \Figref{fig:galvanic}. The independent resonator circuit has a total inductance of $L_r$ and a capacitance to ground $C_r$ at both resonator nodes. The fluxonium circuit consists of a Josephson junction, with a characteristic critical current $I_c$ and energy $E_J$, shunted by its self-capacitance $C_q$ and by an inductor implemented using a linear array of larger Josephson junctions. The total inductance consists of two sections: one section of inductance $L_q$ independent from the resonator circuit, and a smaller section of inductance $L_c$ shared between the resonator and fluxonium which leads to their currents being coupled. This mutual inductive coupling will lead to a magnetic dipole interaction between the resonator field and the phase difference across the qubit Josephson junction. A static magnetic flux $\Phi_{\mathrm{ext}}$ is externally applied to the fluxonium loop.

The equations of motion for the node fluxes are given by Kirchoff's law of current conservation at each node
\begin{widetext}
\begin{align}\label{nodeEq}
\frac{2}{L_r} \left( \Phi_R - \Phi_b \right) + C_r \ddot{\Phi}_R &= 0 \qquad \mathrm{(node\,\,R)}\nonumber\\
\frac{2}{L_r} \left( \Phi_L - \Phi_a \right) + C_r \ddot{\Phi}_L &= 0 \qquad \mathrm{(node\,\,L)}\nonumber\\
\frac{2}{L_q} \left( \Phi_2 - \Phi_b \right) + I_c\sin \frac{2\pi}{\Phi_0}\left( \Phi_2 - \Phi_1 - \Phi_{\text{ext}}\right)  + C_q\left( \ddot{\Phi}_2 - \ddot{\Phi}_1 \right) &= 0 \qquad \mathrm{(node\,\,1)}\nonumber\\
\frac{2}{L_q} \left( \Phi_1 - \Phi_a \right) - I_c\sin \frac{2\pi}{\Phi_0}\left( \Phi_2 - \Phi_1 - \Phi_{\text{ext}}\right)  - C_q\left( \ddot{\Phi}_2 - \ddot{\Phi}_1 \right) &= 0 \qquad \mathrm{(node\,\,2)}\nonumber\\
\frac{2}{L_r} \left( \Phi_R - \Phi_b \right) - \frac{1}{L_c}\left( \Phi_b - \Phi_a \right) - \frac{2}{L_q}\left( \Phi_b - \Phi_2 \right) &= 0 \qquad \mathrm{(node\,\,a)}\nonumber\\
\frac{2}{L_r} \left( \Phi_L - \Phi_a \right) + \frac{1}{L_c}\left( \Phi_b - \Phi_a \right) - \frac{2}{L_q}\left( \Phi_a - \Phi_1 \right) &= 0 \qquad \mathrm{(node\,\,b)}.
\end{align}
\end{widetext}

Similar to the analysis for the resonator chain, we define the new variables $\Phi_r^{\pm} \triangleq \Phi_R \pm \Phi_L$, $\Phi_q^{\pm} \triangleq \Phi_2 \pm \Phi_1$, $\Phi_s^{\pm} \triangleq \Phi_b \pm \Phi_a$. The equations of motion for the differential variables are obtained from Eqs.~\ref{nodeEq}
\begin{equation}
\begin{aligned}\label{diffEq}
\frac{2}{L_r} \left( \Phi_r^{-} - \Phi_s^{-}\right) + C_r \ddot{\Phi}_r^{-} &= 0\\
\frac{2}{L_q} \left( \Phi_q^{-} - \Phi_s^{-}\right) + 2I_c\sin\frac{2\pi}{\Phi_0}\left( \Phi_q^{-} - \Phi_{\mathrm{ext}} \right) + 2C_q \ddot{\Phi}_q^{-} &= 0\\
\frac{2}{L_r} \left( \Phi_r^{-} - \Phi_s^{-}\right) - \frac{2}{L_c}\Phi_s^{-} -\frac{2}{L_q}\left(\Phi_s^{-} - \Phi_q^{-}\right) &= 0.
\end{aligned}
\end{equation}
and we can express the shunt differential variable in terms of the resonator and fluxonium differential variables
\begin{align}\label{phis}
\Phi_s^{-} = \frac{L_q L_c}{L_\Sigma^2} \Phi_r^{-} + \frac{L_r L_c}{L_\Sigma^2} \Phi_q^{-},
\end{align}
where for brevity we define $L_\Sigma^2 \triangleq L_r L_q + L_q L_c + L_c L_r$.

The fluxonium and resonator have the same center of mass $\Phi_q^{+} = \Phi_r^{+} = \Phi_s^{+}$ (see Eqs.~\ref{nodeEq}), as expected from the symmetry of the circuit. Additionally, the COM variables are not coupled to the differential ones, in particular, the junction phase difference $\Phi_q^{-}$ which captures the internal fluxonium states. Replacing the shunt variable (Eq.~\ref{phis}) in Eq.~\ref{diffEq} gives a set of equations just for the resonator and fluxonium
\begin{equation}
\begin{aligned}
\frac{L_q + L_c}{L_\Sigma^2} \Phi_r^{-} + \frac{C_r}{2} \ddot{\Phi}_r^{-} - \frac{L_c}{L_\Sigma^2}\Phi_q^{-} &= 0\\
\frac{L_r + L_c}{L_\Sigma^2} \Phi_q^{-} + I_c\sin\frac{2\pi}{\Phi_0}\left(\Phi_q^{-} - \Phi_{\mathrm{ext}}\right) + C_q \ddot{\Phi}_q^{-} - \frac{L_c}{L_\Sigma^2}\Phi_r^{-} &= 0.
\end{aligned}
\end{equation}

The Euler-Lagrange equations of motion can be directly derived from the following Lagrangian
\begin{equation}
\begin{aligned}
{\mathcal{L}} &= \frac{1}{2}\left(\frac{C_r}{2}\right) {{\dot{\Phi}}_r}^{- 2} - \frac{1}{2}\frac{L_q + L_c}{L_\Sigma^2}{\Phi^{-}_r}^{2} + \frac{L_c}{L_\Sigma^2} \Phi_r^{-} \Phi_q^{-}\\
&+ \frac{1}{2}C_q{{\dot{\Phi}}_q}^{- 2} -\frac{1}{2}\frac{L_r + L_c}{L_\Sigma^2}{{\Phi}^{-}_q}^{2} + E_J\cos\frac{2\pi}{\Phi_0}\left(\Phi_q^{-} - \Phi_{\mathrm{ext}} \right).
\end{aligned}
\end{equation}

The canonical conjugate momenta, corresponding to the charge variables, are given by $Q_r^{-} = \partial \mathcal{L}/\partial {\dot{\Phi}_r}^{-} = (C_r/2) {\dot{\Phi}_r}^{-}$ for the resonator and $Q_q^{-} = \partial \mathcal{L}/\partial {\dot{\Phi}_q}^{-} = C_q {\dot{\Phi}_q}^{-}$ for the fluxonium.

Following a Legendre transformation, the circuit Hamiltonian reads
\begin{equation}
\begin{aligned}
\mathrm{H} & = Q_r^{-} {\dot{\Phi}_r}^{-} + Q_q^{-} {\dot{\Phi}_q}^{-} - \mathcal{L}\\
& = \frac{{Q^{-}_r}^{2}}{C_r} + \frac{{\Phi^{-}_r}^{2}}{2L_r^\prime}+ \frac{L_c}{L_\Sigma^2} \Phi_r^{-} \Phi_q^{-}\\
& + \frac{{Q^{-}_q}^{2}}{2 C_q} + \frac{{\Phi^{-}_q}^{2}}{2L_r^\prime} - E_J \cos\frac{2\pi}{\Phi_0}\left( \Phi^{-}_q - \Phi_{\mathrm{ext}} \right)
\end{aligned}
\end{equation}

Moving to the quantum picture, the flux and charge variables are promoted to quantum operators, ${\op{\Phi}}_n$ and ${\op{Q}}_n$, which obey the canonical commutation relation for bosonic operators $\left[{\op{\Phi}}_{n}, {\op{Q}}_{m}\right] = \mathrm{i}\hbar\delta_{n,m}$. For brevity, the minus superscript is removed since all variables are differential. The total Hamiltonian can be decomposed as $\mathbf{H} = \mathbf{H}_{\mathrm{r}} + \mathbf{H}_{\mathrm{q}} + \mathbf{H}_{\mathrm{int}}$ with separate terms corresponding to the resonator, fluxonium and fluxonium-resonator dipole interaction, respectively.

The resonator Hamiltonian can be written in a second quantized form as
\begin{align}
\mathbf{H}_{\mathrm{r}} = \frac{{{\op{Q}}_r}^{2}}{C_r} + \frac{{\op{\Phi}}_r^{2}}{2L_r^\prime}  = \hbar\omega_r\left(\op{a}^\dagger \op{a} +\frac{1}{2}\right),
\end{align}
writing the charge and flux operators in terms of raising (lowering) operators $\op{a}^\dagger$ ($\op{a}$), ${\op{\Phi}}_r = \sqrt{\hbar Z_r/2} \left(\op{a}^\dagger  + \op{a}\right)$ and ${\op{Q}}_r = \mathrm{i}\sqrt{\hbar/2 Z_r}\left(\op{a}^\dagger  - \op{a}\right)$. The cavity resonance frequency and impedance is given by $\omega_r = \sqrt{2/L^\prime_r C_r}$ and $Z_r = \sqrt{2L^\prime_r/C_r}$. Due to the galvanic coupling with the fluxonium circuit, the renormalized resonator inductance becomes $L^\prime_r = L_\Sigma^2/(L_q + L_c) = L_r + (L_q \parallel L_c)$, equivalent to a contribution from the parallel combination of the bare fluxonium inductance and coupling inductance.

Moving to the fluxonium Hamiltonian, we can write it in the familiar form \cite{Manucharyan_Science2009}
\begin{align}
\mathbf{H}_{\mathrm{q}} &= 4E_C {\op{n}}_q^2 + \frac{1}{2} E_L {\op{\varphi}}_q^2  - E_J \cos\left({\op{\varphi}}_q - \varphi_{\mathrm{ext}} \right) \\
&= \sum_l \varepsilon_l | l \rangle\langle l |,
\end{align}
using the charge number ${\op{n}}_q = {\op{Q}}_q/2e $ and phase ${\op{\varphi}}_q = 2\pi {\op{\Phi}}_q / \Phi_0$ operators. The charging and inductive energies are defined as $E_C = e^2/2C_q$ and $E_L = \left(\Phi_0/2\pi \right)^2/L^\prime_q$, and the flux bias phase is defined as $\varphi_{\mathrm{ext}} = 2\pi\Phi_{\mathrm{ext}}/\Phi_0$. Similarly, due to the galvanic coupling, the renormalized fluxonium inductance becomes $L^\prime_q = L_\Sigma^2/(L_r + L_c) = L_q + (L_r \parallel L_c)$, equivalent to a contribution from the parallel combination of the bare resonator inductance and coupling inductance. To calculate the fluxonium eigenspectrum, the phase difference across the junction is discretized in a 1D grid and the operators are expressed in matrix form in this grid basis. The eigenvalues $\varepsilon_l$ and eigenstates $|l\rangle$ are then obtained from diagonalizing the fluxonium Hamiltonian in the grid basis.

\begin{figure}[!t]
	\begin{center}
		\includegraphics[width=0.5\textwidth]{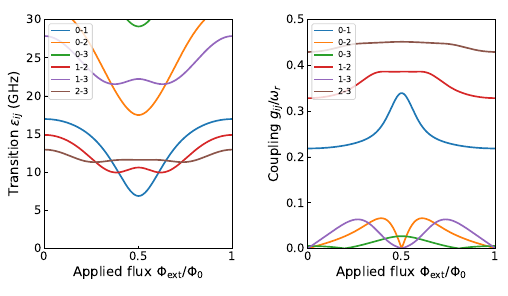}
	\end{center}
	\vspace{-0.5cm}
	\caption{\label{fig:fluxonCoup}\textbf{Fluxonium-resonator coupling strength}. (left) Transition frequencies between the ground state and first three excited states of the fluxonium circuit and (right) their normalized coupling strengths with the edge-resonator as a function of applied flux.} 
\end{figure}

The magnetic dipole coupling is given by the term
\begin{equation}
\begin{aligned}\label{Hgalvanic}
\mathbf{H}_{\mathrm{int}} &= \frac{L_c}{L_\Sigma^2} {\op{\Phi}}_r  \cdot {\op{\Phi}}_q\\
&= \frac{\omega_r}{Z_r} \frac{L_c}{L_c + L_q} \sqrt{\frac{\hbar Z_r}{2}} \left(\op{a}^\dagger + \op{a} \right) \cdot \frac{\Phi_0}{2\pi} \sum_{l,l^\prime} \langle l| {\op{\varphi}}_q |l^\prime \rangle |l\rangle \langle l^\prime| \\
&= \sum_{l,l^\prime} \hbar g_{l,l^\prime} \op{\sigma}_{l,l^\prime} \left(\op{a}^\dagger + \op{a} \right),
\end{aligned}
\end{equation}
where we expanded the qubit flux operator in terms of all possible transitions between fluxonium eigenstates $l \rightarrow l^\prime$ coupled to the cavity field through the dipole matrix elements $\langle l| {\op{\varphi}}_q |l^\prime \rangle$. 
These matrix elements dictate the selection rules for the fluxonium circuit and allow non-trivial transitions between eigenstates differing by more than one quantum number, unlike the transmon. The coupling amplitude between the oscillator current and the fluxonium dipole, normalized by the oscillator resonance frequency, can be rewritten as
\begin{equation}
\begin{aligned}\label{normG}
\frac{g_{l,l^\prime}}{\omega_r} = \frac{\Phi_0}{2\pi} \frac{L_c}{L_c + L_q} \left(2 \hbar Z_r\right)^{-\frac{1}{2}} \,\, \langle l| {\op{\varphi}}_q |l^\prime \rangle .
\end{aligned}
\end{equation}
This form emphasizes how the normalized coupling strength depends on the resonator impedance, fluxonium matrix element, and the inductive participation which sets the fraction of the qubit phase bias ${\op{\varphi}}_q$ interacting with the resonator. The full circuit parameters and range of coupling inductances (outlined in Table \ref{table:CircParams}) are chosen to reach the USC regime $g_{0,1}/\omega_r \approx 1$ (see \Figref{fig:fluxonCoup}).

\subsubsection{Tunable coupling}

Following the expression in Eq.~\ref{normG}, the normalized fluxonium-resonator coupling can be varied by tuning the coupling inductor $L_c$. Operating in the parameter regime $L_q > L_c \gg L_r$, the effective fluxonium shunting inductance $L_q^\prime =  L_q + (L_r \parallel L_c) \approx L_q$, and thus $E_L$, does not significantly change with $L_c$. Given that the other parameters $E_J$ and $E_C$ remain fixed, this approach is suitable for maintaining the same fluxonium energy spectrum while varying the normalized light-matter coupling. 

The coupling inductor is implemented as a chain of flux-tunable asymmetric SQuIDs as shown in the diagram in Fig.~\ref{fig:1}b. Each SQuID is defined as a ring interrupted by two junctions with different energies, $E_{J1}$ and $E_{J2}$, with a relative asymmetry defined as $d  = (E_{J2} - E_{J1})/(E_{J2} + E_{J1})$. The SQuID Hamiltonian is given by $\mathbf{H}_J = -E_{J1}\cos{\op{\varphi}}_1 -E_{J2}\cos{\op{\varphi}}_2$, where $\op{\varphi}_1, \op{\varphi}_2$ represent the phase difference across each junction. For an externally applied magnetic flux $\Phi_{\mathrm{ext}}$, these phase differences satisfy the fluxoid quantization condition in the loop $\op{\varphi}_2 - \op{\varphi}_1 + 2\pi \Phi_{\mathrm{ext}}/\Phi_0 = 2\pi m$, where $m\in\mathbb{Z}$. Defining the phase difference across the SQuID as $\op{\varphi} = (\op{\varphi}_2 + \op{\varphi}_1)/2$, the Hamiltonian can be rewritten as a single cosine potential with a tunable junction energy~\citep{Koch_PRA2007}
\begin{equation}
\begin{aligned}\label{SQuID}
\mathbf{H}_{J} &= -(E_{J1} + E_{J2})\cos\left(\pi\frac{\Phi_{\mathrm{ext}}}{\Phi_0}\right)\\
&\sqrt{1 + d^2 \tan^2 \left(\pi\frac{\Phi_{\mathrm{ext}}}{\Phi_0}\right)} \cos(\hat{\op{\varphi}} - \varphi_0)\\
&\triangleq -E_J^\prime (\Phi_{\mathrm{ext}}) \cos(\hat{\op{\varphi}} - \varphi_0),
\end{aligned}
\end{equation}
where the phase shift in the potential $\varphi_0 = \tan^{-1}[d \tan(\pi\Phi_{\mathrm{ext}}/\Phi_0)]$ can be disregarded by a change of variables. The SQuID can thus be regarded as a flux-tunable inductor $L_J^\prime = \left(\Phi_0/2\pi\right)^2/E_J^\prime$. The coupling element is implemented as $M=4$ SQuIDs connected in series with a total inductance $L_c = M L_J^\prime$. Since the SQuID junctions have a similar size as the junctions used for the fluxonium inductor, we can treat the SQuID array as a tunable linear inductor and neglect the nonlinearities originating from the cosine potential.

\begin{figure}[!t]
	\begin{center}
		\includegraphics[width=0.5\textwidth]{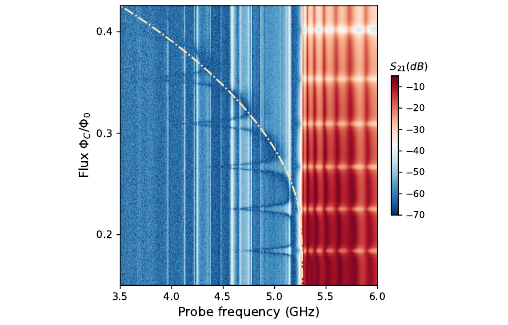}
	\end{center}
	\vspace{-0.5cm}
	\caption{\label{fig:LcCal}\textbf{Coupling inductor calibration}. Transmission spectrum near the lower band edge. The frequency shift of the lowest lattice mode due to inductance renormalization is fitted (dotted line) to estimate the values of the SQuID coupling inductors.} 
\end{figure}

\subsubsection{Coupling to a resonator chain}

After describing the linear resonator chain in Eq.~\ref{Htb} and the inductive coupling between a resonator and a qubit in Eq.~\ref{Hgalvanic}, it becomes straightforward to combine these to describe the full Hamiltonian where the fluxonium is coupled to the multi-mode metamaterial
\begin{equation}
\begin{aligned}\label{Hqb}
\mathbf{H}/\hbar = &\sum_j \omega_j \op{a}_j^\dagger \op{a}_j 
+ \sum_l \varepsilon_l | l \rangle\langle l | \\
- &\sum_{\langle i,j \rangle} t_{ij}\left(\op{a}_i^\dagger\op{a}_j + \op{a}_j^\dagger \op{a}_i^\dagger\right) + \sum_{l,l^\prime} g_{ll^\prime} \op{\sigma}_{ll^\prime} \left(\op{a}_0^\dagger + \op{a}_0 \right),
\end{aligned}
\end{equation}
where the fluxonium is coupled to the edge resonator (site 0) in the tight-binding chain. The expression for the on-site cavity resonances $\omega_j$ and photon hopping rates $t_{ij}$ follow the relations in Eq.~\ref{termsTB}, with modifications from the inductive coupling element at the edge and finite size of the chain. It is assumed that the fluxonium circuit and the SQuID coupler do not modify the capacitance matrix $\hat{\op{C}}$ for the resonator chain.

The resonator sites in the bulk have the same resonance frequency and impedance, given by $\omega_r = L_r^{-\frac{1}{2}} \left[\hat{\op{C}}^{-1}_\pm\right]_{jj}^{\frac{1}{2}}$ and $Z_r = L_r^{\frac{1}{2}}\left[\hat{\op{C}}^{-1}_\pm \right]_{jj}^{\frac{1}{2}}$. 
This makes the hopping rate between the bulk sites $t = t_{ij} = - \frac{1}{2} Z_r \left[\hat{\op{C}}^{-1}_\pm\right]_{i,j}$. 
For the edge site not coupled to the qubit ($j = N$), the diagonal and off-diagonal capacitance matrix elements are different from the bulk since the the edge is coupled on side to another resonator and on the other to the $50\,\Omega$ waveguide. This difference in the capacitive loading leads to a very small change in the on-site resonance $\omega_N\simeq\omega_r$ and hopping rate $t_{N, N-1}\simeq t$, which is nevertheless taken into consideration in the numerical analysis. 
In addition to a capacitive perturbation, the edge site coupled to the qubit ($j=0$) has also an inductive perturbation. The edge resonator inductance has contributions from both the coupler and qubit $L^\prime_r = L_r + (L_q \parallel L_c)$, which is mostly given by the coupler inductance for the choice of device parameters. This leads to modified expressions for the resonance frequency $\omega_r^{\prime} = L_r^{\prime\frac{1}{2}}\left[\hat{\op{C}}^{-1}_\pm \right]_{1,1}^{\frac{1}{2}}$, impedance $Z_r^\prime = L_r^{\prime \frac{1}{2}}\left[\hat{\op{C}}^{-1}_\pm \right]_{1,1}^{\frac{1}{2}}$ and hopping $t^{\prime} = - \frac{1}{2} \sqrt{Z_r Z^{\prime}_r} \left[\hat{\op{C}}^{-1}_\pm\right]_{1,2}$. 
This inductive perturbation leads to a lattice mode being shifted outside the single-particle band. We reduce the bare inductance of this edge resonator such that this perturbation in the resonance frequency does not decouple the edge resonator from the rest of the lattice. Additionally, the frequency shift in the lowest eigenmode is used to extract the SQuID inductor values (see \Figref{fig:LcCal}).

\subsection{Qubit Stark shift calculation}\label{sec:stark_calib}

The qubit frequency shift in Fig.~\ref{fig:4}b is attributed to an ac-Stark shift induced by pumping the lowest lattice mode, having a resonant frequency $\omega_{k=0}$ and linewidth $\kappa$, with a coherent drive of power $P_d$. We use this measurement to provide an estimate for the steady state average photon population $\bar{n} = P_d/\hbar \omega_{k=0} \kappa$ in the driven mode. The line attenuation in the experimental setup has been calibrated to evaluate the input drive power $P_d$ at the device. The Stark shift is given by $\chi_{k=0, l=1} \bar{n}$, where $\chi_{k,l}$ corresponds to the dispersive shift between the lattice mode $k$ and fluxonium state $|l\rangle$. This dispersive shift was calculated following the procedure outlined in~\cite{Zhu_PRB2013} $\chi_{k,l} = \sum_{l^\prime} (\chi_{k, ll^\prime} - \chi_{k, l^\prime l})$. The partial dispersive shifts between mode $k$ and the $l \rightarrow l^\prime$ transitions are given by $\chi_{k, ll^\prime} = |g_{k, ll^\prime}|^2/(\varepsilon_{ll^\prime} - \omega_k)$, with the dipole coupling $g_{k, ll^\prime} = u_{k,0} g_{ll^\prime}$ being rescaled by the eigenmode participation $u_{k,0}$ at the $j=0$ edge lattice site.

\section{Elastic scattering}

\begin{figure*}[t]
	\begin{center}
		\includegraphics[width=1\textwidth]{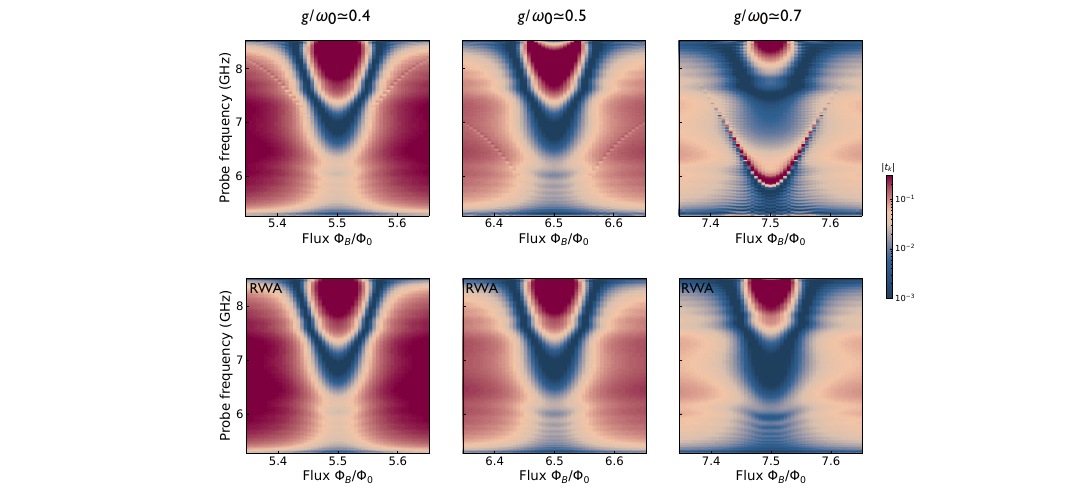}
	\end{center}
	\vspace{-0.5cm}
	\caption{\label{fig:MPS}\textbf{MPS simulations}. Normalized elastic scattering coefficient using MPS, for the coupling strengths $g/\omega_0$ probed in the experiment (see top labels). The upper plots represent the full model, while the lower plots represent the RWA model without the counter-rotating terms.} 
\end{figure*}

For the theoretical treatment of the elastic scattering, we describe the device using an infinite tight-binding model coupled locally to a fluxonium, captured by the Hamiltonian 
\begin{equation}
\begin{aligned}\label{eq:H_scat}
\mathbf{H}/\hbar = \omega_0&\sum_{j=-\infty}^{\infty} \op{a}_j^\dagger \op{a}_j -\delta \op{a}_0^\dagger\op{a}_0
+ \sum_l \varepsilon_l | l \rangle\langle l | \\
- t &\sum_{i} \left(\op{a}_i^\dagger \op{a}_{i+1}+\text{H.c}\right) + \sum_{l,l^\prime} g_{ll^\prime} \op{\sigma}_{ll^\prime} \left(\op{a}_0^\dagger + \op{a}_0 \right),
\end{aligned}
\end{equation}
where the fluxonium is coupled to the cavity site $j=0$. We take into account the multi-level structure of the fluxonium, as well as the detuning, $\delta$, of the $j=0$ cavity frequency due to the coupling inductor. 

\subsection{Scattering coefficients for a two-level system within the rotating-wave approximation}\label{sec:SCAT-RWA-TLS}

Truncating the fluxonium to the two lowest energy states and performing the rotating-wave approximation (RWA), we can derive simple analytical expressions for the scattering coefficients of a single photon. 
Writing a single-particle eigenstate as 
\begin{equation}
\begin{aligned}
&|\psi_k \rangle = \left[ \sum_x u_{k}(x)\op{a}_x^\dagger+c_{1}\op{\sigma}_+\right]|0, 0_{\mathbf{k}}\rangle,\\
&u_k(x)=\begin{cases}
e^{ikx}+R_ke^{-ikx},\quad{x<0},\\
T_ke^{ikx},\quad {0<x},
\end{cases}
\end{aligned}
\end{equation}
where $|0, 0_{\mathbf{k}}\rangle$ corresponds to the ground state of the fluxonium and the vacuum of all the photonic modes.
We can solve for the scattering coefficients $R,T$, giving
\begin{equation}
\label{eq:scat-coefs-RWA}
\begin{split}
T_k =& \frac{2t\sin(k)}{-\delta+G(k)-2ti\sin(k)},\\
R_k =& -\frac{\delta+G(k)}{-\delta+G(k)-2ti\sin(k)},
\end{split}
\end{equation}
where $G(k)=g^2/(\omega_k-\Delta)$ and $\omega_k = \omega_0-2t\cos(k)$. We have also defined $\Delta = \varepsilon_1$ and $g=g_{10}$ for simplicity. 

\Cref{eq:scat-coefs-RWA} admits two resonance-like behaviors. First, when $\Delta= \omega_k$, i.e., the incoming photon is resonant with the qubit transition, we have that $T\rightarrow 0,R\rightarrow 1$, as expected and as was predicted for a uniform cavity array \cite{Zhou2008a}.
Second, when $G(k)-\delta=0$, i.e., $\omega_k = \Delta+g^2/\delta$, we have $R\rightarrow 0,T\rightarrow 1$. Thus, we see that the detuned cavity (with detuning $\delta$) acts as an impurity, giving rise to a non-zero reflection even in the absence of the fluxonium. 

\subsection{Matrix-product-states }\label{sec:MPS}
To go beyond the RWA and the two-level truncation of the fluxonium, we employ matrix-product-states (MPS), using the methods of Refs. \cite{Belyansky_PRR2021,Sanchez_PRL2014}. We describe the system in real space, using $200$ cavities on each side of the fluxonium. This allows us to simulate the scattering of a single photon free of boundary effects. 
The on-site bosonic Hilbert space is truncated to $5$ Fock states, and all results were found to have converged at this value.
To obtain the elastic scattering coefficients we perform the procedure from \cite{Sanchez_PRL2014}.
We first find the ground state $|GS\rangle$ of the Hamiltonian in \cref{eq:H_scat} using the density-matrix-renormalization-group method. We then create a single-photon Gaussian wavepacket on top of the ground state, i.e., we create the state $|\psi(t=0)\rangle =  \sum_x c_x \op{a}_x^\dagger|GS\rangle$, where $c_x =\mathcal{N}e^{-\frac{(x-x_0)^2}{2\sigma^2}+ik_0x}$, and $\mathcal{N}$ is a normalization so that $\sum_x |c_x|^2=1$. We choose the wavepacket parameters so that it spans the whole frequency band, and so that it starts localized on one side of the spin.
We then evolve this state for a sufficiently long time $t_\infty$, until the scattering process has ended. From the resulting state, $|\psi(t_\infty)\rangle$, we extract the overlaps $\langle GS|\op{a}_{x}|\psi(t_\infty)\rangle$ for all $x$. Fourier transforming this quantity, and dividing by $c_k$ (the Fourier transform of the wavepacket $c_x$) gives us the normalized elastic scattering coefficients, shown in Fig.~\ref{fig:3}c of the main text and reproduced here in the top row of \cref{fig:MPS}.
In the bottom row of \cref{fig:MPS} we show the same simulations under the RWA, i.e we consider \cref{eq:H_scat} but with the counter-rotating terms ($\sum_{l>l^\prime}(\op{\sigma}_{l l^\prime}\op{a}^\dagger +\text{h.c.})$) removed. This is not equivalent to \cref{sec:SCAT-RWA-TLS} since here we include higher levels of the fluxonium and their coupling to the cavity.
In particular, $g_{02},g_{12}$ couple the $N=1,2,3$ number of excitations sectors, even in the absence of the counter-rotating terms.
Nevertheless, for our device parameters, $g_{02}\ll g_{01},\varepsilon_1$, and as a result, the Fano resonances that are visible in the top row of \cref{fig:MPS} are essentially invisible in the RWA model in the bottom row of \cref{fig:MPS}.
The theory curves in Fig.~\ref{fig:3} are calculated from diagnolizing the tight-binding Hamiltonian in \cref{Hqb} with flux-tunable parameters. Given the large Hilbert space of the joint system, the Hamiltonian was truncated to a maximum number of excitations. The discrepancy between the calculated energy levels for the two- and three-photon bound states and the MPS resonances is likely caused by this truncation restriction.

\section{Semi-classical calculation of the nonlinear emission}\label{sec:MM_meanfield}

We use input-output theory~\cite{clerkRMP_2010} to describe the nonlinear emission investigated in Section~\ref{sec:MMentanglement}. The Langevin equations of motion for the cavity modes are given by
\begin{equation}
\partial_t \op{}a_j = -i[\op{H},\op{a}_j]-\frac{\kappa_j}{2}\op{a}_j+\sqrt{\kappa}\op{a}_{\mathrm{in},j},
\end{equation}
where $\kappa_j$ is the decay rate of cavity $j$ and $\op{a}_{in,j}$ is the input field operator for cavity $j$. We assume only the first and last cavities are coupled to external environments (i.e., $\kappa_j=0$ except for $\kappa_1=\kappa_N\equiv \kappa$ where $N$ is the number of cavities). The first cavity is being coherently driven with a frequency $\omega_d$ and effective amplitude $\varepsilon_d$, i.e., $\langle \op{a}_{\mathrm{in},1}(t)\rangle =\varepsilon_d\cos(\omega_dt)/\sqrt{\kappa}$. The output field from the last cavity, corresponding to the heterodyne measurement performed in the experiment, is given by the input-output relation
\begin{equation}
\label{eq:input-output}
\langle \op{a}_{\mathrm{out}}(t)\rangle = \sqrt{\kappa}\langle \op{a}_{N}(t)\rangle,
\end{equation}
and its Fourier transform $\langle \op{a}_{\mathrm{out}}(\omega)\rangle$ provides the measured output field at a probe frequency $\omega$.

A brute-force computation of \cref{eq:input-output} is not feasible due to the strong drive, and consequently, the large Hilbert space explored.
Instead, we proceed with a semi-classical approximation, assuming a mean-field ansatz for the density matrix $\rho(t) = \bigotimes_j \ket{\alpha_j(t)}\bra{\alpha_j(t)}\otimes \rho_{\text{q}}(t)$, where $\alpha_j$ are coherent states of amplitudes $\alpha_j$, and $\rho_{\text{q}}$ is the fluxonium density matrix truncated to three levels.
\begin{widetext}
The resulting equations of motion for the full model described in \cref{eq:H_scat} are 
\begin{align}
	\partial_t\alpha_i &= -i\omega_i\alpha_i +it(\alpha_{i+1}+\alpha_{i-1})-i\delta_{i,1}\sum_{\ell\ell'}g_{\ell\ell'}\sigma_{\ell,\ell'}-i\delta_{i,1}\epsilon_d\cos(\omega_dt)-\frac{\kappa_i}{2}\alpha_i,\\
	\partial_t\sigma_{m,n} &= -i(\varepsilon_n-\varepsilon_m)\sigma_{m,n}-i(\alpha_1+\alpha_1^*)\left(\sum_{\ell'}g_{n\ell'}\sigma_{m,\ell'}-\sum_{\ell}g_{\ell m}\sigma_{\ell,n}\right),
\end{align}
where $\sigma_{i,j}\equiv \langle j|\rho_{\text{q}}|i\rangle$ and $i,j =0,1,2$.

Under the RWA, the equations of motions are instead 
\begin{align}
	\partial_t\alpha_i &= -i\omega_i\alpha_i +it(\alpha_{i+1}+\alpha_{i-1})-i\delta_{i,1}\sum_{\ell<\ell'}g_{\ell\ell'}\sigma_{\ell,\ell'}-i\delta_{i,1}\epsilon_d\cos(\omega_dt)-\frac{\kappa_i}{2}\alpha_i.\\
	\partial_t\sigma_{m,n} &= -i(\varepsilon_n-\varepsilon_m)\sigma_{m,n}-i\alpha_1\left(\sum_{\ell'<n}g_{n\ell'}\sigma_{m,\ell'}-\sum_{\ell>m}g_{\ell m}\sigma_{\ell,n}\right)-i\alpha_1^*\left(\sum_{\ell'>n}g_{n\ell'}\sigma_{m,\ell'}-\sum_{\ell<m}g_{\ell m}\sigma_{\ell,n}\right).
\end{align}
\end{widetext}

\begin{figure}[!t]
	\begin{center}
        \includegraphics[width=0.5\textwidth]{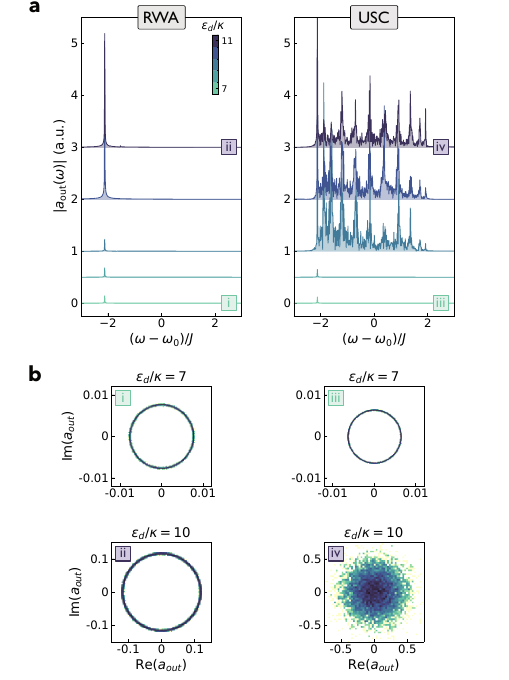}
	\end{center}
    \vspace{-0.5cm}
    \caption{\label{fig:MM_meanfield}\textbf{Multimode mean-field calculations.} Numerically investigating the response of the system from the semi-classical equations of motion for the RWA and full (USC) model. The output field from the last cavity is analyzed by looking at (\textbf{a}) the magnitude of its Fourier transform and (\textbf{b}) its trajectory in phase space. 
    } 
\end{figure}

These coupled equations are solved for a relatively large number of cavities, $N = 10$, and the Fourier transform of the solution $\alpha_N(t)$ corresponds to the output field plotted in the main text in \Figref{fig:4}. To capture the steady-state response, the initial transient behavior (cavity ring-up) is removed before taking the Fourier transform. The numerical results are summarized in \Figref{fig:MM_meanfield}, where we investigate the output field when we drive the input cavity at the lowest eigenmode frequency and a range of coherent drive strengths, for the equation describing the RWA and the full model. We also solved the coupled equations for the $N=26$ case (corresponding to the total number of cavities used in the experiment), and while we found qualitatively similar results to the $N=10$ scenario, we proceeded with solving for different drive amplitudes for the latter case due to faster convergence of numerics.

In the RWA model, we recover a linear response, where the output cavity field oscillates at the drive frequency and follows a circular trajectory in phase space. As expected, the emission amplitude at $\omega = \omega_d$ increases with the drive amplitude.

In the full model, including the counter-rotating coupling terms, we recover a nonlinear multimode response similar to the one observed in \Figref{fig:4}. Above a certain threshold drive amplitude, the output spectrum displays resonances not only confined to the driven mode but extends to encompass all the additional dressed eigenmodes of the lattice. In contrast to the linear response, the output field displays a complex and chaotic behavior, where the phase space trajectory no longer converges to a periodic orbit, but instead explores a large region in phase space. For the parameters explored in these mean-field calculations, we can thus far conclude the importance of the fluxonium multi-level structure and ultrastrong coupling for observing the nonlinear multimode dynamics.

\section{Multimode correlations and entanglement}\label{sec:MM_HZ}

For characterizing the inelastically emitted microwave fields we adopt the formalism in Refs.~\cite{Eichler_PRL2011,Eichler_PRA2012} for probing the quantum state of propagating microwave photons and their correlations using linear amplifiers and quadrature amplitude detectors. In the case of a single mode, the output field $\op{a}$ goes through a phase insensitive amplifier of gain $G$ which introduces an additional noise mode $\op{h}$. The amplified field is then down-converted in a microwave mixer using a local oscillator (LO), in order to detect the in-phase and quadrature components $\op{I}$ and $\op{Q}$. These quantities are related through the complex amplitude operator defined as $\op{S} = \sqrt{G}(\op{a} + \op{h}^\dagger) = \op{I} + i\op{Q}$.

For probing correlations in the emitted microwave fields, the device output is split at room temperature into two separate quadrature detection branches. We perform homodyne measurements, where each branch has a separate local oscillator tuned in frequency to match the desired mode frequencies $\omega_\alpha$ and $\omega_\beta$, respectively. The time domain traces of the complex field amplitudes $\op{S}_{\alpha,\beta}$ on each branch are recorded simultaneously, and the gain prefactors are accounted from calibrated values at the mode frequencies $G_\alpha = G(\omega_\alpha)$ and $G_\beta = G(\omega_\beta)$, respectively. The details on the microwave hardware (filters, amplifiers, ADC measurement bandwidth) used for these correlation measurements are outlined in \cref{Sec:expsetup}.

Under the assumption that the noise added by the detection chain is uncorrelated with the generated signal, the field moments can be decomposed into products of signal and noise moments:
\begin{equation}
\begin{aligned}\label{SaSb}
\langle (\op{S}_\alpha^\dagger)^n \op{S}_\beta^m\rangle = G_\alpha^{\frac{n}{2}} G_\beta^{\frac{m}{2}}\sum_{i,j = 0}^{n,m} \binom{n}{i}\binom{m}{j} \\
\langle (\op{a}_\alpha^\dagger)^i \op{a}_\beta^j\rangle \langle (\op{h}_\alpha^{n-i} (\op{h}_\beta^\dagger)^{m-j}\rangle.
\end{aligned}
\end{equation}

For the measured inelastic multi-mode emission presented in Fig.~\ref{fig:4}a, the power spectrum $\langle \op{a}^\dagger_\omega \op{a}_\omega\rangle$ is extracted from the measured moments of the signal $\op{S}_\omega$ and noise $\op{h}_\omega$, at a single homodyne branch, by sweeping the local oscillator frequency around the single-photon band. Processing the time domain traces such that the noise amplitude has zero mean $\langle\op{h}^\dagger\rangle = 0$ yields the simplified relationship between signal and noise power terms:
\begin{equation}
\begin{aligned}\label{SS}
\langle \op{S}_\omega^\dagger \op{S}_\omega\rangle = G(\omega) \left( \langle \op{a}_\omega^\dagger \op{a}_\omega\rangle + \langle \op{h}_\omega \op{h}_\omega^\dagger\rangle\right).
\end{aligned}
\end{equation}
Therefore, the power spectrum is measured from the signal power subtracted by the noise power. The noise power term is extracted from the output noise of the detection chain when the waveguide is not driven by the pump tone.

\subsection{Hillery-Zubairy criteria}

\begin{figure*}[!t]
	\begin{center}
		\includegraphics[width=0.88\textwidth]{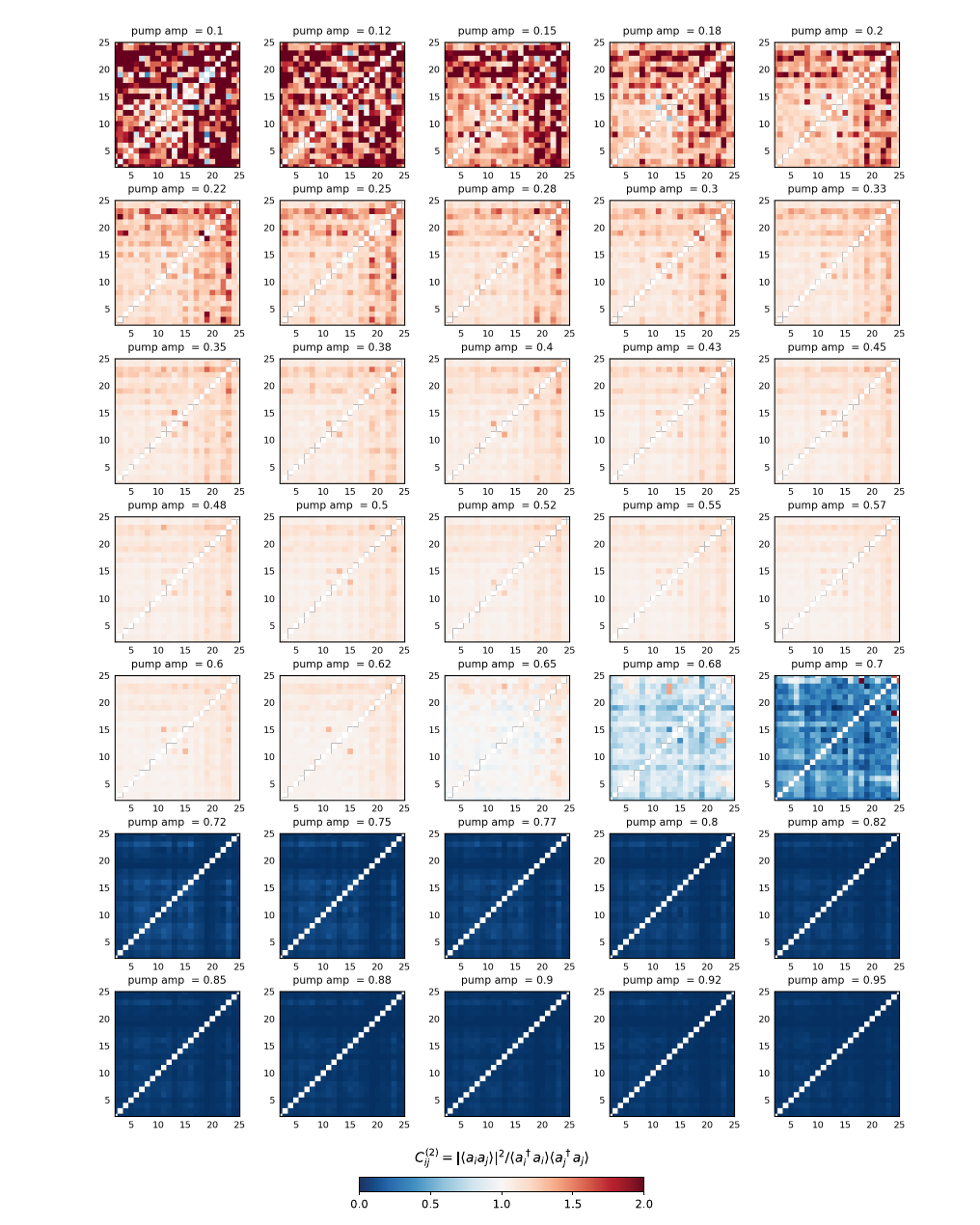}
	\end{center}
	\vspace{-0.5cm}
	\caption{\label{fig:HZMaps}\textbf{Hillery-Zubairy criterion}. Two-mode squeezing correlators $C_{ij}$ as a function of the pump amplitude. The diagonal elements $i = j$ are not a relevant measure of entanglement and have been removed.} 
\end{figure*}

To characterize the entanglement properties of the generated multimode state we rely on the Hillery-Zubairy criterion for two-mode states \cite{Hillery_Zubairy_PRL2006}. For two harmonic modes with annihilation operators $\op{a}_A$ and $\op{a}_B$, respectively, the following Cauchy-Schwarz inequality holds for pure product states.

\begin{equation}
\begin{aligned}
|\langle\op{a}_A \op{a}_B\rangle|^2 = |\langle \op{a}_A\rangle|^2|\langle \op{a}_B\rangle|^2 \leq \langle \op{a}^\dagger_A \op{a}_A\rangle \langle \op{a}^\dagger_B \op{a}_B\rangle .
\end{aligned}
\end{equation}
Furthermore, Hillery and Zubairy have shown that this inequality holds for any separable state, generalised as a mixture of pure product states $\rho^{AB} = \sum p_n |\Psi_n^{AB}\rangle\langle \Psi_n^{AB}|$. A violation of this inequality implies that the two-mode state is entangled, which is a necessary but not sufficient condition since there exist two-mode entangled states that satisfy this inequality.

This entanglement criterion justifies our choice of two-mode correlators $\mathcal{C}_{ij} = |\langle \op{a}_i \op{a}_j\rangle|^2/\langle \op{a}^\dagger_i \op{a}_i\rangle \langle \op{a}^\dagger_j \op{a}_j\rangle$, measured for every pair of waveguide modes. The photon intensity $\langle \op{a}^\dagger_i \op{a}_i\rangle$ is measured from the power spectrum, while the squeezing correlators $\langle \op{a}_i\op{a}_j\rangle$ are extracted from the dual homodyne measurement of the second order moments:
\begin{equation}
\begin{aligned}\label{SaSb}
\langle \op{S}_i \op{S}_j\rangle = \sqrt{G_i G_j} \left( \langle \op{a}_i \op{a}_j\rangle + \langle \op{h}^\dagger_i \op{h}_j^\dagger\rangle\right).
\end{aligned}
\end{equation}
It is important to emphasize that the expression for the correlator $\mathcal{C}_{ij}$ makes it insensitive to uncertainties in the estimated values of the gain prefactors, $G_i$ and $G_j$, since these cancel out from evaluating the ratio of the second order moments.
The map of squeezing correlators for every pair of waveguide modes, and its dependence on the pump amplitude, is displayed in \Figref{fig:HZMaps}.

\section{Commercial Disclaimer}
Specific product citations are for clarification only and are not an endorsement by the authors or NIST.

\newpage
\clearpage
\bibliography{main}

\end{document}